\documentclass[11pt,urlcolor=blue, linkcolor=blue]{article} 
\usepackage{cite}

\usepackage{amsmath, amsthm, amssymb,slashed}
\usepackage{ifpdf}
\ifpdf
  \usepackage[pdftex]{graphicx}
  \usepackage{epstopdf}
\else
  \usepackage[dvips]{graphicx}
\fi
\parskip=\baselineskip

\usepackage[usenames, dvipsnames]{color}

\allowdisplaybreaks[1]

\newcommand{\cblue}[1]{\textcolor{black}{#1}}
\newcommand{\cred}[1]{\textcolor{black}{#1}}
\newcommand{\cgreen}[1]{\textcolor{black}{#1}}

\definecolor{mygray}{gray}{0.6}

\usepackage{upgreek}
\usepackage{bbm}

\newcommand{\mZ}{\mathbbm{Z}}
\newcommand{\GSD}{\text{GSD}}

\newcommand\finline[3][]{\begin{myfont}[#1]{#2}#3\end{myfont}}%
\newenvironment{myfont}[2][]{\csname#2\endcsname[#1]}{}

\newcommand{\bea}{\begin{eqnarray}}
\newcommand{\eea}{\end{eqnarray}}
\def\be{\begin{equation}}
\def\ee{\end{equation}}

\def\RP{{\mathbb{RP}}}

\usepackage{color}
\usepackage[colorlinks,citecolor=blue]{hyperref}
\definecolor{red}{rgb}{1,0,0}
\definecolor{blue}{rgb}{0,0,1}
\definecolor{dblue}{rgb}{0,0,0.4}
\definecolor{green}{rgb}{0,1,0}
\definecolor{black}{rgb}{0,0,0}
\definecolor{white}{rgb}{1,1,1}

\definecolor{brn}{rgb}{.8,.4,.0}
\definecolor{redo}{rgb}{1,.5,.0}
\definecolor{ddgrn}{rgb}{0,0.4,0}
\definecolor{dgrn}{rgb}{0,0.55,0}
\definecolor{dbl}{rgb}{0,0,0.5}

\usepackage[bbgreekl]{mathbbol}
\usepackage{amscd}

\newcommand{\Z}{\mathbb{Z}}

\newcommand{\R}{\mathbb{R}}

\newcommand{\ii}{i}

\newcommand{\Tr}{{\rm Tr}}

\newcommand{\pt}{\partial}

\newcommand{\bpm}{\begin{pmatrix}}
\newcommand{\epm}{\end{pmatrix}}
\newcommand{\bmm}{\begin{matrix}}
\newcommand{\emm}{\end{matrix}}

\newcommand{\cH}{ {\cal H} }

\newcommand{\cS}{ {\cal S} } 
\newcommand{\cT}{ {\cal T} }

\def\CD{{\cal D}}

\def\CF{{\cal F}}

\def\CH{{\cal H}}

\def\CL{{\cal L}}

\def\CO{{\cal O}}

\def\CS{{\cal S}}
\def\CT{{\cal T}}

\def\CV{{\cal V}}

\def\Z{{\mathbb{Z}}}

\def\R{{\mathbb{R}}}

\def\pt{\mathrm{pt}}
\def\Tr{{\mathrm{Tr}}}

\begin{document}
\begin{titlepage}
\begin{flushright}
\end{flushright}
\vskip 1.25in
\begin{center}


{\bf\LARGE{Braiding Statistics and 
Link Invariants of\\[2.75mm]
Bosonic/Fermionic 
Topological Quantum Matter\\[3.75mm] 
in 2+1 and 3+1 dimensions 
}}

\vskip0.5cm 
\Large{  Pavel Putrov$^{1}$, Juven Wang$^{1,2,3}$, 
and Shing-Tung Yau$^{4,2,3}$} 
\vskip.5cm
 {\small{\textit{$^1$School of Natural Sciences, Institute for Advanced Study, Princeton, NJ 08540, USA}\\}}
 \vskip.2cm
 {\small{\textit{$^2${Center of Mathematical Sciences and Applications, Harvard University,  Cambridge, MA, USA} \\}}
}
 \vskip.2cm
{\small{\textit{$^3${Department of Physics, Harvard University,  Cambridge, MA 02138, USA} \\}}
}
 \vskip.2cm
{\small{\textit{$^4$Department of Mathematics, Harvard University, Cambridge, MA 02138, USA \\}}
}

\end{center}
\vskip.5cm
\baselineskip 16pt
\begin{abstract}

Topological Quantum Field Theories (TQFTs) 
pertinent to some emergent low energy phenomena of condensed matter lattice models
in 2+1 and 3+1 dimensions are explored. 
Many of our TQFTs are highly-interacting without free quadratic analogs.
Some of our bosonic TQFTs can be regarded as the continuum field theory formulation of Dijkgraaf-Witten twisted discrete gauge theories.
Other bosonic TQFTs beyond the Dijkgraaf-Witten description and all fermionic TQFTs (namely the spin TQFTs)
are either higher-form gauge theories where particles must have strings attached,
or
fermionic discrete gauge theories obtained by gauging the 
fermionic Symmetry-Protected Topological states (SPTs).
We analytically calculate both the Abelian and non-Abelian braiding statistics data of anyonic particle and string excitations in these theories, where
the statistics data can one-to-one characterize the underlying topological orders of TQFTs.
Namely, we derive path integral expectation values of links formed by line and surface operators in these TQFTs.
The acquired link invariants include not only the familiar Aharonov-Bohm linking number,
but also Milnor triple linking number in 3 dimensions, 
triple and quadruple linking numbers of surfaces,
and intersection number of surfaces in 4 dimensions.
We also construct new spin TQFTs with the corresponding knot/link invariants of Arf(-Brown-Kervaire), Sato-Levine and others. 
We propose a new relation between the fermionic 
SPT partition function and the Rokhlin invariant.
As an example, we can use these invariants and other physical observables, including ground state degeneracy, 
reduced modular $\mathcal{S}^{xy}$ and $\mathcal{T}^{xy}$ matrices,
and the partition function on $\mathbb{RP}^3$ manifold,
to identify all $\nu \in \mathbbm{Z}_8$ classes of 2+1 dimensional gauged
$\mathbb{Z}_2$-Ising-symmetric $\mathbb{Z}_2^f$-fermionic Topological Superconductors 
(realized by stacking $\nu$ layers of a pair of chiral and anti-chiral $p$-wave superconductors [$p+ip$ and $p-ip$], where boundary supports non-chiral Majorana-Weyl modes)
with continuum spin-TQFTs.


\end{abstract}
\end{titlepage}

\tableofcontents   

\newpage

\section{Introduction and Summary}

In condensed matter physics, 
we aim to formulate a systematic framework within unified principles to understand 
many-body quantum systems and their underlying universal phenomena.
Two strategies are often being used: classification and characterization.
The classification aims to organize the distinct macroscopic states / phases / orders of quantum matter in terms of distinct classes, give these classes some proper 
mathematical labels,
and find the mathematical relations between distinct classes. 
The characterization aims to distinguish different classes of matter in terms of some universal physics probes as
incontrovertible experimental evidence of their existences. 
Ginzburg-Landau theory \cite{Landau1937,GL5064,Landau1958} provides a framework to understand the global-symmetry breaking states and
their phase transitions. Ginzburg-Landau theory uses the group theory in mathematics to classify the states of matter through their global symmetry groups.
Following Ginzburg-Landau theory and its refinement to the Wilson's renormalization-group theory \cite{WilsonKogut1974}, 
it is now well-known that we can characterize symmetry breaking states through 
their gapless Nambu-Goldstone modes, the long-range order (see References therein \cite{anderson1984basic}),  
 and their behaviors through the critical exponents. In this classic paradigm,
physicists focus on looking into the long-range correlation function of \emph{local} operators $\CO(x)$ at a spacetime point $x$, or into a generic $n$-point correlation function:
\bea
\langle \CO(x_1) \CO(x_2) \rangle,  \;\;\;\;\;\; \langle \CO_1(x_1) \CO_2 (x_2) \cdots  \CO_n (x_n)\rangle,  \;\;\;\;\;\; etc.
\eea
through its long-distance behavior. 

However, a new paradigm beyond-Ginzburg-Landau-Wilson's have emerged since the last three decades \cite{WenBook, sachdev2011quantum}.
One important theme is the 
emergent conformal symmetries and emergent gauge fields at the quantum 
critical points of the phase transitions. 
This concerns the critical behavior of gapless phases of matter where the energy gap closes to zero at the infinite system size limit. 
Another important theme is the \emph{intrinsic topological order} \cite{Wen:1990tm}.
The topological order cannot be detected through the local operator $\CO(x)$, 
nor the Ginzburg-Landau symmetry breaking order parameter,
nor the long-range order. 
Topological order is famous for harboring fractionalized anyon excitations that have the fractionalized statistical Berry phase\cite{Wilczek:1990ik}.
Topological order should be characterized and detected through the 
\emph{extended} or \emph{non-local}  operators.
It should be classified through the quantum pattern of the long-range entanglement (See \cite{Wen1610.03911} for a recent review).
Topological order can occur in both gapless or gapped phases of matter.
In many cases, 
when topological orders occurr in the gapped phases of condensed matter system,
they may have low energy effective field theory descriptions by Topological Quantum Field Theories (TQFTs) \cite{Witten:1988hf}.
Our work mainly concerns gapped phases of matter with intrinsic topological order that have TQFT descriptions.

One simplest example of topological order in 2+1 dimensions 
(denoted as 2+1D\footnote{
We denote 
$n+1$ dimensional spacetime as $n+1$D, 
with $n$ dimensional space and 1 dimensional time.
})
is called the $\Z_2$ topological order\cite{WenPRBZ2TO44.2664}, equivalently the $\Z_2$ spin liquid\cite{ReadSachdevPRL66.1773}, 
or the $\Z_2$ toric code \cite{Kitaev2003}, or the $\Z_2$ discrete gauge theory \cite{Wegner:1971jf}.
Indeed the $\Z_2$ topological order exists in any dimension, 
say in $((d-1)+1)$D with $d\geq 3$. Discrete 
$\Z_N$ gauge theory can be described by an integer level-$N$ $BF$ field theory with an action $\int\,\frac{N}{2\pi}B\wedge dA$,
where $A$ and $B$ are locally 1-form and $d-2$-form gauge fields.
The case of $N=2$ and $d=3$ is our example of 2+1D $\Z_2$ topological order.
Since $n$-point correlation function of local operators cannot detect the nontrivial $\Z_2$ or $\Z_N$ topological order, 
we shall instead use \emph{extended} operators to detect its nontrivial order.
The \emph{extended} operators pf  are Wilson and 't Hooft operators:
$W_{A,e_n}({C_n^1}) =\exp[\ii e_n \oint_{C_n^1} A ]$ 
carrying the electric charge $e_n$
along  a closed curve ${C_n^1}$, 
and  $W_{B,q_m}({S_m^{d-2}}) =   \exp[\ii  q_m \oint_{S_m^{d-2}}B ]$ 
carrying the magnetic charge $q_m$
along a closed surface ${S_m^{d-2}}$.
The path integral  (details examined in the warm up exercise done in Section \ref{sec:BdA}) 
for the correlator of the extended operators results in
\bea
\langle W_{A,e_n}({C_n^1}) W_{B,q_m}({S_m^{d-2}})   \rangle = \exp[ -\ii  \frac{2\pi}{N}  e_n q_m \text{Lk}(C_n^1, S_m^{d-2})].
\eea
With some suitable
$e_n$ and $q_m$ values, its expectation value is nontrivial (i.e. equal to 1), if and only if
the linking number $\text{Lk}(C_n^1, S_m^{d-2})$ of the line and surface operator is nonzero. 
The closed line operator $W_{A,e_n}({C_n^1})$ can be viewed as creating and then annihilating a pair of $e_n$ particle-antiparticle 0D anyon excitations
along a 1D trajectory in the spacetime.
The closed surface operator $W_{B,q_m}({S_m^{d-2}})$ can be viewed as creating and then annihilating 
a pair of $q_m$ fractionalized flux-anti-flux $(d-3)$D excitations 
along some trajectory in the spacetime
(Note that the flux excitation is a 0D anyon particle in 2+1D, while it is a 1D anyonic string excitation in 3+1D).
A nontrivial linking implies that there is a nontrivial braiding process between $e_n$ charge and $q_m$ flux excitation in the spacetime\footnote{
Let us elaborate on what exactly is meant by this. Let $\CL$ be a link in a closed space-time manifold $M$. The link $\CL$ can be decomposed as some submanifolds, including lines or surfaces. 
The lines or surfaces become operators in TQFT that create anyonic excitations at their ends. For example, an open line creates the anyonic particle at two end points. An open surface creates the anyonic string at its boundary components. A closed line thus creates a pair of anyonic particle/anti-particle from vacuum, and then annihilate them to vacuum. The closed surface creates the anyonic strings from vacuum and then annihilate them to vacuum. Therefore the link $\CL$ can be viewed as the time trajectory for the braiding process of those anyonic excitations, 
where braiding process means the time-dependent process (a local time as a tangent vector in a local patch of the whole manifold) that is moving those anyonic excitations around to form a closed trajectory as the link of submanifolds (lines, surfaces) in the spacetime manifold. 
The braiding statistics concerns the complex number that arise in the path integral with the configuration described above. The braiding statistics captures the statistical Berry phase of excitations of particle and string.
We remark \emph{quantum dimensions} of anyonic particles/strings in Sec.~\ref{sec:conclude}.
}.
The link confugurations shown in terms of spacetime braiding process are listed in Table \ref{table:TQFTlink}.
Physically, 
we can characterize the topological order through the statistical Berry phase between anyonic excitations, say $\exp[ \ii  \frac{2\pi}{N}  e_n q_m]$, via
the nontrivial link invariant.
Mathematically, the viewpoint is the opposite,
the topological order, or TQFT, or here the $BF$  theory detects the nontrivial link invariant. 
It shall be profound to utilize both viewpoints to explore the topological order in condensed matter, TQFT in field theories, and 
link invariants in mathematics.
This thinking outlines the deep relations between quantum statistics and spacetime topology\cite{JWangthesis, 1602.05951}.

The goals of our paper are: 
(1) Provide concrete examples of topological orders and TQFTs that occur in 
 emergent low energy phenomena 
in some well-defined fully-regularized many-body quantum systems. 
(2) Explicit exact analytic calculation of the braiding statistics and link invariants for our topological orders and TQFTs.
For the sake of our convenience and for the universality of low energy physics, we shall approach our goal through TQFT, without worrying about a particular lattice-regularization or
the lattice Hamiltonian.
However, we emphasize again that all our TQFTs are low energy physics of some well-motivated lattice quantum Hamiltonian systems,
and we certainly shall either provide or refer to the examples of such lattice models and condensed matter systems, cases by cases.
To summarize, our TQFTs / topological orders shall satisfy the following physics properties:
\begin{enumerate}
\item The system is unitary. 
\item Anomaly-free in its own dimensions. Emergent as the infrared low energy physics from 
fully-regularized microscopic many-body quantum Hamiltonian systems with a ultraviolet high-energy lattice cutoff. This motivates a practical purpose for condensed matter.
\item The energy spectrum has a finite energy gap $\Delta$ in a closed manifold for the microscopic many-body quantum Hamiltonian systems.
We shall take the large energy gap limit $\Delta \gg 1$ to obtain a valid TQFT description. 
The system can have degenerate ground states (or called the zero modes) on a closed spatial manifold $M^{d-1}$. 
This can be evaluated 
as the path integral on the manifold $M^{d-1} \times S^1$, namely
$Z(M^{d-1} \times S^1)= \dim \cH_{M^{d-1}} \equiv \GSD$
as the dimension of Hilbert space, which counts the ground state degeneracy (\GSD).  
On an open manifold, the system has the lower dimensional boundary theory 
with anomalies.
The anomalous boundary theory could be gapless. 
\item The microscopic 
Hamiltonian contains the short-ranged local interactions between the spatial sites or links. 
The Hamiltonian operator is Hermitian.
Both the TQFT and the Hamiltonian system are defined within the local Hilbert space.
\item The system has the long-range entanglement, and contains fractionalized anyonic particles, anyonic strings, or other extended object as excitations. 
\end{enumerate}

As said, the $\Z_2$ topological order / gauge theory 
 has both TQFT and lattice Hamiltonian descriptions \cite{WenPRBZ2TO44.2664, ReadSachdevPRL66.1773, Kitaev2003, Wegner:1971jf}.
There are further large classes of topological orders, including the $\Z_2$ toric code \cite{Kitaev2003},
 that can be described by 
a local short-range interacting Hamiltonian:
\be\label{eq:Hamiltonian}
  \hat{H}=-\sum_v \hat{A}_v-\sum_f \hat{B}_f,
\ee
where $\hat{A}_v$ and $\hat{B}_f$ are mutually commuting bosonic lattice operators acting on the vertex $v$ and the face $f$ of a triangulated/regularized space.
With certain appropriate choices of $\hat{A}_v$ and $\hat{B}_f$, we can write down an exact solvable spatial-lattice model 
(e.g. see a systematic analysis in \cite{Wan1211.3695, Wan:2014woa}, and also similar models 
in \cite{MesarosRan, Jiang:2014ksa, Wang1404.7854}) whose low energy physics
yields the Dijkgraaf-Witten topological gauge theories \cite{Wittencohomology}.  
Dijkgraaf-Witten topological gauge theories in $d$-dimensions are defined in terms of path integral on a spacetime lattice ($d$-dimensional manifold ${M^{d}}$  triangulated with $d$-simplices).
The edges of each simplex are assigned with quantum degrees of freedom of a gauge group $G$ with group elements $g \in G$.
 Each simplex then is associated to a complex $U(1)$ phase of $d$-cocycle  $\omega_d$ of the cohomology group $H^{d}(G,U(1))$
 up to a sign of orientation related to the ordering of vertices (called the branching structure). How do we
 convert the spacetime lattice
path integral $Z$
 as the ground state solution of the Hamiltonian given in
 Eq. (\ref{eq:Hamiltonian})?
We design the $\hat{B}_f$ term as the zero flux constraint on each face / plaquette.
 We design that the $\hat{A}_v$ term acts on the wavefunction of a spatial slice through each vertex $v$ by lifting 
 the \cred{initial} state through an imaginary time evolution to a new state with a vertex $v'$ via
 $\hat{A}_v=\frac{1}{|G|}\sum_{[vv']=g\in G}\hat{A}_v^g$. Here the edge along the imaginary time is assigned with $[vv']=g$ and 
 all $g \in G$ are summed over.  The precise value of $\hat{A}_v^g$ is related to fill the imaginary spacetime simplices with cocycles  $\omega_d$.
 The whole term $\hat{A}_v$ can be viewed as the near neighbor interactions that capture the statistical Berry phases
 and the statistical interactions.
 Such models are also named 
the twisted quantum double model \cite{deWildPropitius:1995cf, Wan1211.3695}, or the twisted gauge theories \cite{Wang1404.7854, Wan:2014woa}, due to the fact that 
Dijkgraaf-Witten's group cohomology description requires twisted cocycles.
 
With a well-motivated lattice Hamiltonian, we can ask what is its low energy \cred{continuum} TQFT.
The Dijkgraaf-Witten model should be described by bosonic TQFT, because its definition does not restrict to a spin manifold. 
Another way to understand this bosonic TQFT is the following.
Since $\hat{A}_v$ and $\hat{B}_f$ are bosonic operators in Eq.\ref{eq:Hamiltonian}, we shall term such a Hamiltonian as a bosonic system and bosonic quantum matter. TQFTs for bosonic Hamiltonians are bosonic TQFTs that require no spin structure.
We emphasize that bosonic quantum matter and bosonic TQFTs have only \emph{fundamental bosons} (without any fundamental fermions),
although these bosonic systems can allow excitations of \emph{emergent anyons}, including \emph{emergent fermions}. 
It has been noticed by \cite{deWildPropitius:1995cf, WangSantosWen1403.5256, Wang1403.7437, Kapustin1404.3230, Wang1404.7854, 1405.7689}
that the cocycle in the cohomology group reveals the continuum field theory action (See, in particular, the Tables in \cite{1405.7689}).
A series of work develop along this direction by formulating a continuum field theory description for
Dijkgraaf-Witten topological gauge theories of discrete gauge groups, 
their topological invariants and physical properties
\cite{Kapustin1404.3230, 1405.7689, Gaiotto:2014kfa,
CWangMLevin1412.1781, Gu:2015lfa, Ye1508.05689, 
RyuChenTiwari1509.04266,CWangCHLinMLevin1512.09111,RyuTiwariChen1603.08429,1602.05951, He1608.05393}.
We will follow closely to 
the set-up of \cite{1405.7689, 1602.05951}.  Continuum TQFTs with level-quantizations
are formulated in various dimensions in Tables of \cite{1405.7689}. 
Dynamical TQFTs with well-defined exact gauge transformations to all orders
and their physical observables are organized in terms of path integrals of with linked line and surface operators in Tables of \cite{1602.05951}.
For example, 
we can start by considering the Dijkgraaf-Witten topological gauge theories given by the cohomology group $H^{d}(G,U(1))$,
say of a generic finite Abelian gauge group $G= \prod_I \Z_{N_I}$.
Schematically, leaving the details of level-quantizations into our main text,
in 2+1D, 
we have field theory actions of 
$\int BdA$, 
$\int K_{IJ} A_I dA_J$, 
$\int B_I dA_I+A_I dA_J$ and
$\int B_I dA_I+A_1 A_2A_3$, etc.
In 3+1D, we have
$\int B_I dA_I+A_J A_K dA_L$, 
$\int B_I dA_I+A_1 A_2A_3A_4$.
Here $B$ and $A$ fields are locally 2-form and 1-form gauge fields respectively.
For simplicity, we omit the wedge product ($\wedge$) in the action.
(For example, $A_1 A_2A_3$ is a shorthand notation for $A_1 \wedge A_2 \wedge A_3$.)
The indices of $A_I$ and $B_I$ are associated to the choice of $\Z_{N_I}$ subgroup  in $G= \prod_I \Z_{N_I}$.
The $A$ fields are 1-form U(1) gauge fields, but the $B$ fields can have modified gauge transformations when we turn on the cubic and quartic interactions
in the actions.
We should warn the readers \emph{not} to be confused by the notations:
the TQFT gauge fields $A$ and $B$, and the microscopic Hamiltonian operator $\hat{A}_v$ and $\hat{B}_f$ are totally different subjects.
Although they are mathematically related by the group cohomology cocycles, the precise physical definitions are different.
How do we go beyond the twisted gauge theory description of Dijkgraaf-Witten model?
Other TQFTs that are beyond Dijkgraaf-Witten model, such as 
$\int B_I dA_I+B_IB_J$ \cite{Ye1410.2594, Gaiotto:2014kfa} \cgreen{and other higher form TQFTs\cite{Kapustin:2013uxa}}, may still be captured by the analogous lattice Hamiltonian model
in Eq. (\ref{eq:Hamiltonian}) by modifying the decorated cocycle in $\hat{H}$ to more general cocycles. 
Another possible formulation for beyond-Dijkgraaf-Witten model can be the Walker-Wang model\cite{WalkerWang1104.2632, Williamson:1606.07144}.
The lattice Hamiltonian can still be written in terms of certain version of Eq. (\ref{eq:Hamiltonian}).
All together, 
we organize the list of aforementioned TQFTs,
braiding statistics and link invariants that we compute,
and some representative realizable condensed matter/lattice Hamiltonians, in Table \ref{table:TQFTlink}.

Most TQFTs in the Table \ref{table:TQFTlink} are bosonic TQFTs that require no spin manifold/structure.
However,  $\int \frac{N_I}{2\pi}{B^I  \wedge d A^I} + { \frac{ p_{IJ}}{4 \pi}} A^I \wedge d A^J$ in 2+1D, and $\int \frac{N_I}{2\pi}B^I\wedge dA^I+\frac{p_{IJ}N_IN_J}{4\pi N_{IJ}}\,B^I\wedge B^J$ in 3+1D,
\footnote{
Throughout our article, we denote  ${N_{12}} \equiv  {\gcd(N_{1}, N_{2})}$, in general ${N_{IJ \dots}} \equiv  {\gcd(N_{I}, N_{J},  \dots)}$.
}
are two examples of fermionic TQFTs (or the so-called spin TQFTs) when $p_{II}$ is an odd integer.
A fermionic TQFTs can emerge only from a fermionic Hamiltonian that contains fundamental fermionic operators satisfying the anti-commuting relations {(see e.g. \cite{Gaiotto1505.05856,Bhardwaj1605.01640})}.
We emphasize that the fermionic quantum matter have \emph{fundamental fermions} (also can have fundamental bosons),
although these fermionic systems can allow excitations of other \emph{emergent anyons}. 
Mathematically, TQFTs describing fermionic quantum matter should be tightened to spin TQFTs that require a spin structure 
\cite{BelovMoore2005ze, jenquin2006spin} (see the prior observation of the spin TQFT in \cite{Wittencohomology}).

We shall clarify how we go beyond the approach of \cite{1405.7689, 1602.05951}.
Ref.\cite{1405.7689} mostly focuses on formulating the \emph{probe-field} action and path integral, so that the field variables that are non-dynamical
and do not appear in the path integral measure. Thus Ref.\cite{1405.7689} is suitable for the context of probing the global-symmetry protected states, 
so-called Symmetry Protected Topological states \cite{XieSPT4} (SPTs, see \cite{Wen1610.03911, Senthil1405.4015, Chiu1505.03535} for recent reviews).
Ref.\cite{1602.05951} includes \emph{dynamical gauge fields} into the path integral, that is the field variables which are dynamical
and do appear in the path integral measure. This is suitable for the context for {
Ref.\cite{1602.05951}
observes the relations between the links of submanifolds (e.g. worldlines and worldsheets whose operators
 create anyon excitations of particles and strings) based on the properties of 3-manifolds and
4-manifolds, and then relates the links to the braiding statistics data computed in Dijkgraaf-Witten model \cite{Wang1403.7437, Wang1404.7854,CWangMLevin1412.1781} and in the path integral of TQFTs. }
In this article, we explore from the opposite direction reversing our target. 
We start from the TQFTs as an input (the first sub-block in the first column in Table \ref{table:TQFTlink}), and determine the associated
mathematical link invariants independently (the second sub-block in the first column in Table \ref{table:TQFTlink}). 
We give examples of nontrivial links in 3-sphere $S^3$  and 4-sphere $S^4$, and their path integral expectation value as statistical Berry phases 
(the second column in Table \ref{table:TQFTlink}), and finally associate the related condensed matter models
(the third column in Table \ref{table:TQFTlink}).


In Table \ref{table:TQFTlink}, we systematically survey various link invariants together with relevant braiding processes (for which the invariant is a nontrivial number as $1$) that 
either are new to or had occurred in the literature in a unified manner.
The most familiar braiding is the Hopf link
with two linked worldlines of anyons in 2+1D spacetime\cite{{Witten:1988hf},Wilczek:1990ik} such that $\text{Lk}(\gamma_I,\gamma_J)=1$ .
The more general Aharonov-Bohm braiding \cite{Aharonov1959} or the charge-flux braiding 
has a worldline of an electric-charged particle linked with a ${(d-2)}$-worldsheet of a magnetic flux linked with 
the linking number $\text{Lk}(S_m^{d-2}, C_n^1)=1$ in $(d-1)$+1D spacetime.
The Borromean rings braiding is useful for detecting certain non-Abelian anyon systems\cite{CWangMLevin1412.1781}.
The link of two pairs of surfaces as the loop-loop braiding (or two string braiding) process is mentioned in 
\cite{AlfordPreskill1992,Bodecker2004PRL,Baez0603085}.
The link of three surfaces as the three-loop braiding (or three string braiding) process is 
discovered in 
\cite{Wang1403.7437,Jiang:2014ksa} and explored in \cite{Wang1404.7854}.
The link of four 2-surfaces as the four-loop braiding (or four string braiding) process is explored in 
\cite{CWangMLevin1412.1781,1602.05951,RyuTiwariChen1603.08429}.

More broadly, below we should make further remarks on 
the related work \cite{Kapustin1404.3230, 1405.7689, Ye1410.2594, Gaiotto:2014kfa,
CWangMLevin1412.1781, Gu:2015lfa, Ye1508.05689, 
RyuChenTiwari1509.04266,CWangCHLinMLevin1512.09111,RyuTiwariChen1603.08429, He1608.05393, NingLiuPengYe1609.00985, Ye1610.08645}.
This shall connect our work to other condensed matter and field theory literature in a more general context.
While Ref. \cite{Kapustin1404.3230} is motivated by the discrete anomalies (the 't Hooft anomalies for discrete global symmetries), 
Ref. \cite{1405.7689} is motivated by utilizing locally flat bulk gauge fields as physical probes to detect Symmetry Protected Topological states (SPTs). 
As an aside note, the SPTs are very different from the intrinsic topological orders and the TQFTs that we mentioned earlier:
\begin{itemize}
\item The SPTs are short-range entangled states protected by nontrivial global symmetries of symmetry group $G$.
The SPTs have its path integral $|Z|=1$ on any closed manifold. The famous examples of SPTs include the topological insulators\cite{2010RMP_HasanKane, 2011_RMP_Qi_Zhang} protected by 
time-reversal and charge conjugation symmetries. 
{
The gapless boundaries of SPTs are gappable by breaking the symmetry or introducing strong interactions.
Consequently, take the 1+1D boundary of 2+1D SPTs as an example, the 1+1D chiral central charge is necessarily (but not sufficiently) $c_-=0$.}
 
\item The intrinsic topological orders are long-range entangled states robust against local perturbations, even without any global symmetry protection.
However, some of topological orders that have a gauge theory description of a gauge group $G$ may be obtained by dynamically gauging the global symmetry $G$  of SPTs 
\cite{LG1202.3120, Ye:2013upa}. 
{The boundary theory for topological orders/TQFTs obtained from gauging SPTs must be gappable as well. 
}
\end{itemize}

{In relation to the lattice Hamiltonian, the SPTs has its Hilbert space and group elements associated to
the vertices on a spatial lattice \cite{XieSPT4}, whereas the corresponding group cohomology implementing the homogeneous cocycle
and the holonomies are trivial for all cycles of closed manifold thus $|Z|=1$.
In contrast, the Eq. (\ref{eq:Hamiltonian}) is suitable for topological order that has its Hilbert space and group elements associated to
the links on a spatial lattice \cite{Wan1211.3695,Wang1404.7854,Wan:2014woa}, whereas its group cohomology implementing the inhomogeneous cocycle and its holonomies are non-trivial for cycles of closed manifold thus $|Z|$ sums over different holonomies. }
%

In relation to the field theory, we expect that the SPTs are described by invertible TQFTs (such as the level $N=1$ in $BF$ theory), a nearly trivial theory, but implemented with nontrivial global symmetries. 
(See \cite{Freed2014} for the discussions for invertible TQFTs, and see the general treatment of global symmetries on TQFTs in \cite{Gaiotto:2014kfa}.) 
In contrast, we expect that
the intrinsic topological orders are described by generic non-invertible TQFTs (e.g.  level $N$ $BF$ theory). 
Since Ref.\cite{1405.7689} implements the nearly flat probed gauge fields, the formalism there could not be the complete
story for the intrinsic topological orders and TQFTs of our current interests.
It is later found that one can view the topological actions in terms of dynamical gauge fields instead of the probed fields, 
by modifying the gauge transformations \cite{Gu:2015lfa, Ye1508.05689}.
%
%
Up until now,
there is  good evidence that
we can view the discrete spacetime Dijkgraaf-Witten model
in terms of some continuum TQFTs (See Tables in \cite{1602.05951, 1405.7689} and our 
Table \ref{table:TQFTlink}).
One of the most important issues for understanding the dynamical TQFT 
is  to compute precisely the path integral $Z$ and to
find explicitly the physical observables. 
To this end, one partial goal for this article, is to explicitly compute the path integral and 
the braiding statistics / 
link invariants for these TQFTs in various dimensions. 
We focus mainly on 2+1D and 3+1D for the sake of realistic dimensions in condensed matter physics, but our formalism can be easily applied to any dimension.

Other than TQFTs and discrete gauge theories in Table \ref{table:TQFTlink}, 
we can obtain even more fermionic spin TQFTs by gauging the global symmetries of fermionic SPTs (fSPTs).
An interesting example is gauging the 
fSPTs with 
${\mathbb {Z}}_2^f\times ({\mathbb {Z}}_2)^n$ symmetry in various dimensions.
We are able to address one interesting puzzle concerning the ${\mathbb {Z}}_2^f\times {\mathbb {Z}}_2$ fSPTs as 
Topological Superconductors
with 8 distinct classes labeled by $\nu \in \mZ_8$
\cgreen{(realized by stacking $\nu$ layers of a pair of chiral and anti-chiral $p$-wave superconductors).}
Although it is known that $\nu=0,4$ gauged fSPTs are bosonic Abelian Chern-Simons (CS) theories for bosonic $\Z_2$ gauge and 
twisted gauge theory (toric code and double-semion models), 
and $\nu=2,6$ gauged fSPTs are fermionic Abelian spin-CS theory for fermionic $\Z_2$  gauge and twisted gauge theory, 
the field theories description for the odd-$\nu$ classes ($\nu=1,3,5,7$) are somewhat mysterious.
In some sense, the odd-$\nu$ class are fermionic ``${\mathbb {Z}}_2$ gauge spin-TQFTs,'' but the statistics is somehow non-Abelian.
We solve the puzzle by deriving explicit non-Abelian spin TQFTs obtained from gauging fSPTs, and compute physical observables to distinguish $\nu \in \mZ_8$ class in
Sec. \ref{sec:fTQFT}.

\subsection{The plan of the article and the convention of notation}

%
%

\begin{table}[!h]
\centering
\finline[\fontsize{10}{10}]{Alpine}{
\noindent
\makebox[\textwidth][c] 
{
\begin{tabular}{ccc} 
\hline
$\begin{matrix}\text{(i). TQFT actions}\\[1mm] 
\hline
\hline\\[-4mm]
\text{associated link invariants}
\end{matrix}$  & 
$\begin{matrix}
\text{(ii). Spacetime-braiding process,}\\ 
\text{Path-integral $Z(\text{Link})$/$Z[S^d]$}\\[1mm]
\hline
\hline\\[-4mm]
\text{Quantum statistic braiding data $e^{\ii \theta}$} 
\end{matrix}$ & 
$\begin{matrix}\text{(iii). Comments:}\\ 
\text{Condensed matter models} 
\end{matrix}$ \\
\hline\\[-4mm]
\multicolumn{3}{c}{Any dimensions} \\
\cline{1-3}\\[-2mm]
$
 \begin{matrix}
\text{Sec. \ref{sec:BdA}}: \int \frac{ N_I}{2\pi}{B^I  d A^I} 
\\[2mm]
\hline
\hline\\[-2mm]
 \text{(Aharonov-Bohm) linking number}\\[2mm]
 \text{Lk}(S_m^{d-2}, C_n^1)
 \end{matrix}
$ 
& 
$\begin{matrix}
Z \bpm \includegraphics[scale=0.4]{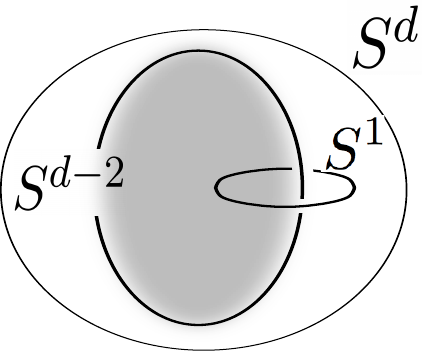}  \epm / Z[S^d]
\\[2mm]
\hline
\hline\\[-2mm]
\exp[ -  \frac{2\pi i}{N} q_m e_n \text{Lk}(S_m^{d-2}, C_n^1)]
\end{matrix}$
&
$\begin{matrix}
\text{$\Z_N$ topological order\cite{WenPRBZ2TO44.2664}}\\
\text{$\Z_N$ spin liquid\cite{ReadSachdevPRL66.1773},}\\ 
\text{$\Z_N$ toric code\cite{Kitaev2003}}\\
\text{$\Z_N$ gauge theory\cite{Wegner:1971jf}} 
\end{matrix}$
 \\
\hline\\[-3mm]
\multicolumn{3}{c}{2+1D} \\
\cline{1-3}\\[-3mm]
$\begin{matrix}
\text{Sec. \ref{sec:AdA}}: \int  \frac{K_{IJ}}{4 \pi} A^I  d A^J,
\\[2mm]
\int \frac{N_I}{2\pi}{B^I   d A^I} + { \frac{ p_{IJ}}{4 \pi}} A^I  d A^J \\[2mm] 
\hline
\hline\\[-2mm]
\text{Linking number: } \text{Lk}(\gamma_I,\gamma_J)
 \end{matrix}
$  
&
$\begin{matrix}
Z \bpm \includegraphics[scale=0.3]{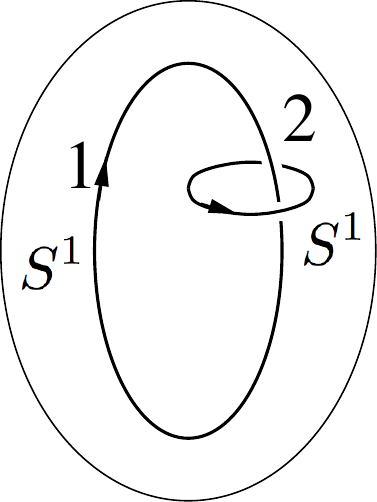} \includegraphics[scale=0.27]{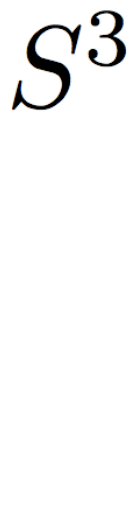}\epm / Z[S^3]  
\\[2mm]
\hline
\hline\\[-2mm]
\exp [- \pi \ii \sum_{I,J} (K^{-1})_{IJ} e_I e_J\text{Lk}(\gamma_I,\gamma_J)]
 \end{matrix}$
&  
$\begin{matrix}
\text{Fractional quantum Hall states\cite{WenKmatrix},}\\
\text{Halperin states\cite{halperin1983theory}}\\
\text{Twisted quantum double\cite{Wan1211.3695, MesarosRan}},\\
\text{String-net models\cite{Levin0404617},}\\
\text{2+1D anyon systems\cite{Kitaev2006, nayak2008non}}\\
\text{({Spin TQFT} for $K_{II}, p_{II} \in$ odd.)}
\end{matrix}$
\\
\hline\\[-2mm]
$\begin{matrix}
\text{Sec. \ref{sec:aaa-theory}}:  \int \frac{ N_I}{2\pi}{B^I   d A^I}+{ \frac{N_1 N_2 N_3\;
p_{}}{{(2 \pi)^2 } N_{123}}} A^1  A^2   A^3 
\\[2mm]
\hline
\hline\\[-2mm]
\text{Milnor's triple linking number}:\\ 
\bar\mu(\gamma_1,\gamma_2,\gamma_3)	
\end{matrix}$  
 & 
 $\begin{matrix}
Z \bpm \includegraphics[scale=0.35]{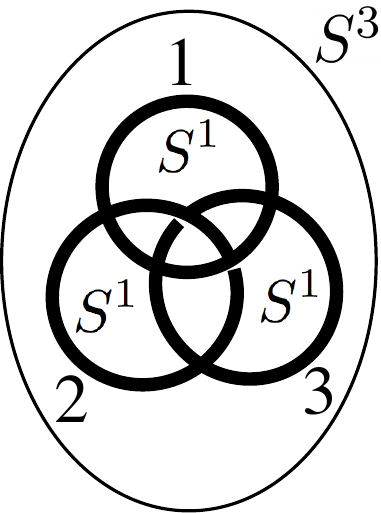} \epm  / Z[S^3]
\\[2mm]
\hline
\hline\\[-2mm]
\exp (
	-\frac{2\pi \ii\,p\,q_1q_2 q_3}{N_{123}}\,\bar{\mu}(\gamma_1,\gamma_2,\gamma_3)
	)
 \end{matrix}$     
& 
$\begin{matrix}
\text{Gauged SPT lattice model\cite{WangSantosWen1403.5256,Gu:2015lfa, He1608.05393}},\\
\text{Twisted quantum double\cite{Wan1211.3695, MesarosRan}},\\
\text{String-net models\cite{Levin0404617},}\\
\text{$D_4$ 
discrete gauge theory \cite{deWildPropitius:1995cf,Wang1404.7854, Gu:2015lfa, He1608.05393}}
\end{matrix}$
\\
\hline
$\begin{matrix}
\text{Sec. \ref{sec:fTQFT}: \cred{Gauged} } \frac{\pi}{4}\int a\cup \text{ABK}
\\[2mm]
\hline
\hline\\[-2mm]
\text{Arf invariant }\\  
 \end{matrix}
$  
&
$\begin{matrix}
Z \bpm \includegraphics[scale=0.35]{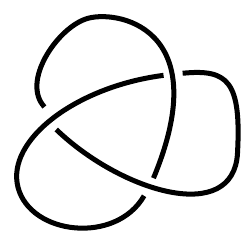} \epm / Z[S^3]  
\\[2mm]
\hline
\hline\\[-2mm]
\pm 1
 \end{matrix}$
&  
$\begin{matrix}
\text{Gauged $\Z_2^f \times \Z_2$-fSPT model.}\\
\text{\cgreen{Odd $\nu \in \mZ_8$ detects knots}}\\
\text{\cgreen{with non-zero Arf invariant (e.g. Trefoil).}}
\end{matrix}$
\\
\hline
$\begin{matrix}
\text{Sec. \ref{sec:fTQFT}: \cred{Gauged} } {\pi}\int a_1\cup a_2 \cup \eta
\\[2mm]
\hline
\hline\\[-2mm]
\text{Sato-Levine invariant }\\  
 \end{matrix}
$  
&
$\begin{matrix}
Z \bpm \includegraphics[scale=0.35]{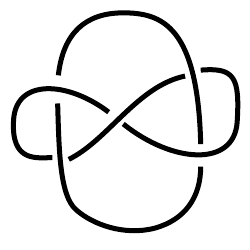} \epm / Z[S^3]  
\\[2mm]
\hline
\hline\\[-2mm]
\pm 1
 \end{matrix}$
&  
$\begin{matrix}
\text{Gauged $\Z_2^f \times (\Z_2)^2$-fSPT model.}\\
\text{\cred{Odd $\nu$ in the mixed class detects}}\\
\text{\cgreen{links with non-zero Sato-Levine}}\\
\text{\cgreen{invariant (e.g. Whitehead).}}
\end{matrix}$
\\
\hline
\\[-2mm]
\multicolumn{3}{c}{3+1D} \\
\cline{1-3}\\[-2mm]
$\begin{matrix}
\text{Sec. \ref{sec:AAdA}}: \int \frac{ N_I}{2\pi}{B^I   d A^I} {{+}}  
\frac{ N_{I'} N_{J'} \; p_{I'J'K'}}{{(2 \pi)^2 } N_{{I'}{J'}}}   
A^{I'}  A^{J'}  d A^{K'} \\[2mm]
%
\hline
\hline\\[-2mm]
\text{Triple linking number of surfaces:}\\
\text{Tlk}(\Sigma_1,\Sigma_3,\Sigma_2)
\end{matrix}$     
&  
$\begin{matrix}
Z \bpm \includegraphics[scale=0.35]{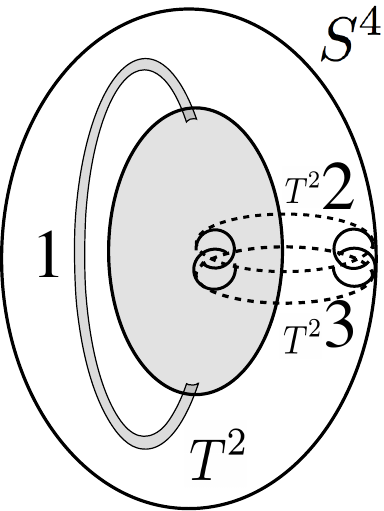} \epm    / Z[S^4]
\\[2mm]
\hline
\hline\\[-2mm]
\exp (
	\frac{2\pi \ii\,p\,q_1 q_2 q_3}{N_{123}}\,\text{Tlk}(\cred{ \Sigma_1,\Sigma_3,\Sigma_2}) 
	)
\end{matrix}$ 
&
$\begin{matrix}
\text{Gauged SPT lattice model\cite{Wang1403.7437}},\\
\text{Twisted gauge theory\cite{Wan:2014woa,ZWang1611.09334}},\\
\text{Abelian string model\cite{Wang1403.7437, Jiang:2014ksa, Wang1404.7854}}\\
\text{}
\end{matrix}$
\\
\hline\\[-2mm]
$\begin{matrix}
\text{Sec. \ref{sec:A4-theory}}: \int \frac{ N_I}{2\pi}{B^I  d A^I} + 
{ \frac{N_1 N_2 N_3 N_4\;
p_{}}{{(2 \pi)^3 } N_{1234}}} A^1  A^2  A^3  A^4 
\\[2mm]
\hline
\hline\\[-2mm]
\text{
Quadruple linking number of surfaces:}\\
\text{Qlk}(\Sigma_1,\Sigma_2,\Sigma_3,\Sigma_4)
\end{matrix}
$
& 
$\begin{matrix}
Z \bpm \includegraphics[scale=0.35]{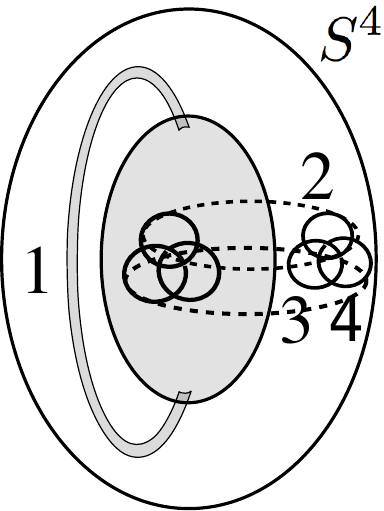} \epm / Z[S^4]
\\[2mm]
\hline
\hline\\[-2mm]
\exp (
	\frac{2\pi \ii\,p\,q_1 q_2 q_3 q_4}{N_{1234}}\,\text{Qlk}(\Sigma_1,\Sigma_2,\Sigma_3,\Sigma_4)
	)
\end{matrix}$ & 
$\begin{matrix}
\text{Gauged SPT lattice model\cite{Gu:2015lfa}},\\
\text{Twisted gauge theory\cite{Wan:2014woa}},\\
\text{Non-Abelian string model\cite{Wang1404.7854}}\\
\text{}
\end{matrix}$
 \\
\hline
$\begin{matrix}
\text{Sec. \ref{sec:BB-theory}}: \int \frac{N_I}{2\pi}B^I dA^I+\frac{p_{IJ}N_IN_J}{4\pi N_{IJ}}\,B^I B^J
\\[2mm]
\hline
\hline\\[-2mm]
\text{Intersection number of surfaces: }\\  
\#(\Sigma_I\cap\Sigma_J)
 \end{matrix}
$  
&
$\begin{matrix}
Z \bpm \includegraphics[scale=0.5]{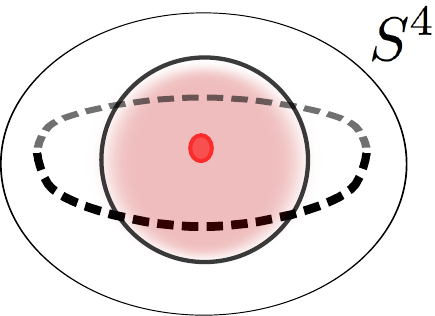} \epm / Z[S^4]  
\\[2mm]
\hline
\hline\\[-2mm]
\exp ( 
	- \frac{ \pi \ii  p_{IJ} e_I e_J}{N_{IJ}}\,\#(\Sigma_I\cap\Sigma_J) )
 \end{matrix}$
&  
$\begin{matrix}
\text{Gauged SPT lattice model\cite{Ye1410.2594}},\\
\text{Walker-Wang like model\cite{WalkerWang1104.2632,KeyserlingkBurnell1405.2988}}.\\
\text{({Spin TQFT} for \cred{$p_{II}, N_{I} \in$ odd}.)}
\end{matrix}$
\\
\hline
\end{tabular}
} \hspace*{35mm}
\caption{TQFT actions, the link invariants computed through the path integral $Z$, and their condensed matter models 
are organized in three columns.
For a comparison of the development from the earlier work, see the setup in \cite{1405.7689, 1602.05951}.
Here $p, p_{IJ}, p_{IJK}$ are quantized integer levels. 
}
\label{table:TQFTlink}
}
\end{table}

The plan of our article is organized as follows.
In Sec.~\ref{sec:BdA}, 
we derive the link invariant of $\int BdA$ theory in any dimension as the Aharonov-Bohm's linking number that detects a charge particle
and a flux loop braiding process through the Aharonov-Bohm phase.
In Sec.~\ref{sec:AdA}, we study $\int K_{IJ} A_I dA_J$ and $\int BdA+AdA$ in 2+1D and show that its path integral calculates the linking number.
In Sec.~\ref{sec:aaa-theory}, we study $\int BdA+A^3$  in 2+1D and obtain Milnor's triple linking number from its path integral.
In Sec.~\ref{sec:AAdA}, we study $\int BdA+A^2dA$ in 3+1D and obtain  triple-linking number of surfaces.
In Sec.~\ref{sec:A4-theory}, we study $\int BdA+A^4$ in 3+1D and obtain quadruple-linking number of surfaces.
In Sec.~\ref{sec:BB-theory}, we study $\int BdA+BB$ in 3+1D and obtain  intersection number of open surfaces.
In Sec.~\ref{sec:fTQFT}, we construct the explicit fermionic SPT path integrals
with ${\mathbb {Z}}_2^f\times ({\mathbb {Z}}_2)^n$ symmetry, and their gauged versions: fermionic spin TQFTs.
We derive the experimentally measurable physics observables, including the ground state degeneracy (GSD), the braiding statistics 
(the modular matrices $\cS^{xy}$ and $\cT^{xy}$), etc. 
In addition, we discuss their relation to various invariants including Arf(-Brown-Kervaire), Rokhlin, Sato-Levine invariants and more.
In Sec.~\ref{sec:conclude}, we conclude with additional remarks.

\cred{
We should emphasize that the link invariants we derive are powerful and important in various aspects.
(1) A link invariant can detect various possible links in spacetime, or various possible braiding processes (regardless} \cgreen{if} \cred{the braiding process is known or unknown to the literature). 
While in the literature, few specific braiding processes have been investigated (such as the three or four string braiding processes),
we can use our link invariants to identify other braiding processes} \cgreen{ that produce nontrivial values of topological invariants and thus have nontrivial statistical Berry phases.
(2) Our method to derive topological invariants is based on field theory description of TQFTs. In particular, our approach is systematic,
using Poincar\'e duality and intersection theory. Our approach is universal, and our result is more general than what appeared in the literature.
}

\noindent
Note: To denote the cyclic group of order $n$, 
we use $\Z_n$ and $\mZ_n$, which are equivalent mathematically, but have different meanings physically.  
We use $\Z_n$ to denote a symmetry group and a gauge group.
We use the slight different notation $\mZ_n$ to denote the distinct classes in the classification of SPTs/TQFTs or in the cohomology/bordism group.
Notation ${\mathbb {Z}}_2^f$ stands for the fermion parity symmetry.
We denote ${N_{IJ \dots}} \equiv  {\gcd(N_{I}, N_{J},  \dots)}$ and $\mZ_{N_{IJ \dots}} \equiv  \mZ_{\gcd(N_{I}, N_{J},  \dots)}$.
\cred{As usual, notation $M_1 \sqcup M_2$ means the disjoint union between two sets or two manifolds 
$M_1$ and $M_2$. The $M\setminus S$ means relative complement of $S$ in $M$.} 
We use $\cup:H^p({M^{d}},\Z_N)\otimes H^q({M^{d}},\Z_N)\rightarrow H^{p+q}({M^{d}},\Z_N)$ to denote cup-product in cohomology ring.
GSD stands for ground state degeneracy. 
In Table.\ref{table:TQFTlink} and elsewhere,
the repeated indices is normally assumed to have Einstein summation, except that the 
$\frac{ N_{I'} N_{J'} \; p_{I'J'K'}}{{(2 \pi)^2 } N_{{I'}{J'}}}   
A^{I'}  A^{J'}  d A^{K'}$ term where \emph{the prime indices here are fixed} instead of summed over.


\section{$\int BdA$ in any dimension and Aharonov-Bohm's linking number} \label{sec:BdA}

Below we warm up by considering the level-$N$ BF theory with an action $\int \frac{N}{2\pi} BdA$ in any dimension, where $N$ is quantized to be an integer.
{{The study of BF theory in physics dates back to the early work of \cite{Horowitz1989, 1991BlauThompson}.}}
Consider the following action on any closed $d$-manifold ${M^{d}}$:
\begin{equation}
	S[A,B]=\int_{{M^{d}}}\,\frac{N}{2\pi}B\wedge dA
\end{equation}
where $A$ is a 1-form gauge field on $M$ and $B$ is a $(d-2)$-form gauge field on $M$. The partition function or path integral without any additional operator insertion
is
\begin{equation}
Z=\int DA DB \exp[\ii S[A,B]] =\int DA DB \exp[\ii \int_{{M^{d}}}\,\frac{N}{2\pi}B\wedge dA]
\end{equation}

Locally the gauge transformation  is given by:
\bea
		A &\rightarrow& A+dg, \\
		B &\rightarrow& B+ d\eta.
\eea
If ${M^{d}}$ has non-trivial topology, globally $g$ and $\nu$ may have discontinuities such that $dg$ and $d\nu$ are continuous forms representing a cohomology class in $2\pi H^1({M^{d}},\Z)$ and $2\pi H^{d-2}({M^{d}},\Z)$ respectively.

Now for a path integral with insertions, let $\Phi$ be a gauge invariant functional $\Phi(A,B)$  of the fields $A$ and $B$.
The path integral with insertion $\Phi$ can be formally defined as
\bea \label{eq:ZBF}
\langle \Phi \rangle =\frac{1}{Z} \int DA \;DB \;\Phi(A,B)\;  \exp[\ii S[A,B]]=\frac{1}{Z} \int DA \;  DB \;  \Phi(A,B)\;  \exp[\ii \int_{{M^{d}}}\,\frac{N}{2\pi}B\wedge dA].
\eea
Let us note that in the case when ${M^{d}}$ has non-trivial topology, the field $B$ only locally can be understood as a $d-2$ form. Globally, it can be realized as $B=\tilde{B}+\beta$ where $\tilde{B}$ is a globally defined $d-2$ form and $\beta$ is a discontinuous $d-2$-form such that $d\beta$ is a continuous form representing a class in $2\pi H^{d-1}({M^{d}},\Z)$, the flux of the $d-2$ gauge field $B$. So the path integral over $\tilde{B}$ actually means the following
\begin{equation}
 \int DB\ldots \equiv\sum_{[d\beta]\in 2\pi H^{d-1}({M^{d}},\Z)}\int D\tilde B\ldots.
\end{equation} 
Below we evaluate the $\langle \Phi \rangle$ in various scenarios starting from the simplest, almost trivial case and gradually increasing complexity. 
\begin{enumerate}

\item If $\Phi (A)$ is independent of the $B$ field, then the integration over $\tilde{B}$ gives the equation of motion as constraint of $A$, which localizes 
$A$ to be flat $U(1)$ connection. Namely, the curvature is zero $F_A=dA=0$. Furthermore, from Poincar\'e duality $H^{d-1}({M^{d}},\Z)\cong H_1({M^{d}},\Z)$, it follows that the sum over fluxes $\beta$ imposes the following constrains on $A$:
\begin{equation}
 \exp ( iN\int_\gamma A) =1,\qquad \forall \gamma
\end{equation} 
that is, modulo gauge transformations, connection $A$ belongs to $\Z_N$  subset of $U(1)$ flat connections:
\begin{equation}
 [A]\in \text{Hom}(H_1({M^{d}}),\Z_N) \subset \text{Hom}(H_1({M^{d}}),U(1)).
\end{equation} 
Note that from the universal coefficient theorem and the fact that $H_0({M^{d}},\Z)$ is a free group, it follows that $\text{Hom}(H_1({M^{d}}),\Z_N)\cong H^1({M^{d}},\Z_N)$. The path integral then reduces to the following finite sum:

\bea  \label{eq:Phi(A)}
\langle \Phi \rangle =\frac{1}{Z} 
\sum_{[A]\,\in  \text{Hom}(H_1({M^{d}}),\Z_N) }
\;\Phi(A).
\eea
The standard normalization for the partition function $Z$ is as follows:
\begin{equation}
 Z=\frac{1}{N}\sum_{[A]\in  \text{Hom}(H_1({M^{d}}),\Z_N) }1
\end{equation} 
so that $Z=1$ for ${M^{d}}=S^{d-1}\times S^1$.

\item If $\Phi (A,B)$ depends on  $B$ field as follows
\bea \label{eq:Phi(A,B)}
\Phi(A,B)=  \prod_{m} \exp[\ii q_m \int_{S_m^{d-2}}B ] \cdot \Phi_0(A).
\eea
Where $\{S_m^{d-2}\}_{m=1,2,\dots}$ is a family of $d-2$-dimensional hypersurfaces inside the spacetime manifold ${M^{d}}$ and $\Phi_0(A)$ is the insertion that depends only on $A$. Gauge invariance requires $q_m\in \Z$.
One can also rewrite (\ref{eq:Phi(A,B)}) as follows:
\begin{equation}
	\Phi(A,B)=  \prod_{m} \exp[\ii q_m \int_{{M^{4}}}B\wedge   \delta^{\perp}({S_m^{d-2}})] \cdot \Phi_0(A)
\end{equation}
where $\delta^{\perp}({S_m^{d-2}})$ is the 2-form valued delta function distribution supported on ${S_m^{d-2}}$. That is,
\bea
\int_{{M^{d}}} \omega_{d-2} \wedge \delta^{\perp}({S_m^{d-2}}) =\int_{S_m^{d-2}} \omega_{d-2}.
\eea
for any $d-2$ form $\omega_{d-2}$. After integrating out $B$ the path integral Eq. (\ref{eq:ZBF}) localizes to the solutions of the equations of motion with source:
\bea \label{eq:FA}
F_A=dA=-\frac{2 \pi}{N} \sum_m q_m \; \delta^{\perp}({S_m^{d-2}}). 
\eea
This equation implies that
$F_A/2\pi$ is a differential form which represents the class in $H^2({M^{d}},\R)$ Poincar\'e dual to  the class \cred{$\frac{1}{N} \sum_m q_m \; [{S_m^{d-2}}]$}
in homology $H_{d-2}(M,\R)$.
Here and below $[S]$ denotes the homology class of the surface $S$.
Since $\frac{F_A}{2 \pi }$ represents the first Chern class $c_1\in H_1({M^{d}},\Z)$ of the $U(1)$ gauge bundle,
$\frac{1}{N} \sum_m q_m \; [{S_m^{d-2}}]$
must represent an integral homology class.
This gives the constraint on the allowed charge $q_m$ (the magnetic charge),
if some of the classes $[{S_m^{d-2}}] \neq 0$ are nontrivial.

%

\item If $H_1({M^{d}},\Z)=0$, then there is a unique solution to Eq. (\ref{eq:FA}), modulo the gauge
redundancy. The cohomology \cred{$H^1({M^{d}} \setminus (\sqcup_m {S_m^{d-2}}),\Z)$}
is then generated by 1-forms $\mu_1, \dots, \mu_m$ such that
\bea
\int_{C_n^1} \mu_i = \delta_{n,i},
\eea
where $C_n^1$ is a small circle linking ${S_m^{d-2}}$.
Here we denote $M\setminus S$ means the relative complement of $S$ in $M$.
The solution of Eq. (\ref{eq:FA}) then becomes:
\bea
A= -\frac{2 \pi}{N} \sum_m q_m \mu_m.\; 
\eea
One possible choice of forms $\mu_m$ is using 1-form valued delta functions supported on $\CV_m^{d-1}$, Seifert hypersurfaces bounded by ${S_m^{d-2}}$ (i.e. such that $\partial\CV_m^{d-1}={S_m^{d-2}}$ and therefore $d\delta^\perp(\CV_m^{d-1})=\delta^\perp(S_m^{d-2})$):
\begin{equation}
	A=-\frac{2 \pi}{N} \sum_m q_m  \delta^\perp(\CV_m^{d-1}).
\end{equation}

\item
If $\Phi_0(A)$ in Eq. (\ref{eq:Phi(A,B)}) is a product of the Wilson loops around the one-dimensional loops $\{ \gamma_n^1 \}$ separate and disjoint from $\{ {S_m^{d-2}} \}$,
such that
\bea \label{eq:Phi(A,B)}
\Phi_0(A)=  \prod_{n} \exp[\ii e_n \int_{\gamma_n^1} A ]  
\eea
with the electric charge $e_n\in\Z$ associated to each loop, then the path integral with $\Phi(A,B)$ insertion can be evaluated as follows:
\begin{multline} \label{eq:ZBFlink}
\langle \Phi \rangle  
= \frac{1}{Z} \int DA \;DB \;  \exp[i S[A,B]]  \;  \exp[i \sum_{n} e_n \int_{\gamma_n^1} A ]   \exp[i  \sum_{m} q_m \int_{S_m^{d-2}}B ]\\
=\exp[ -  \frac{2\pi i}{N} \sum_{m,n} q_m e_n \int_{{M^{d}}}\delta^\perp(\gamma_n^1)\wedge \delta^\perp(\CV_m^{d-1})]
=\exp[ -  \frac{2\pi i}{N} \sum_{m,n} q_m e_n \text{Lk}(S_m^{d-2}, \gamma_n^1)]
\end{multline}
where the $\text{Lk}(S^{d-2}_m, \gamma^1_n)\equiv \#(\CV^{d-1}_m \cap \gamma^1_n)$ is the linking integer number between the loop $\gamma^1_n$ and the $(d-2)$-dimensional submanifold $S^{d-2}_m$, which by definition is given by counting intersection points in $(\CV^{d-1}_m \cap \gamma^1_n)$ with signs corresponding to orientation.

\end{enumerate}

\section{$\int K_{IJ} A_I dA_J$ and $\int BdA+AdA$ in 2+1D and the linking number}
\label{sec:AdA}

In the 2+1D spacetime, as another warp up exercise, consider the action of $U(1)^s$ Chern-Simons theory with level matrix $K$:
\begin{equation}
 S[A]=\int_{{M^{3}}}\sum_{I,J=1}^s \frac{K_{IJ}}{4\pi}A^I\wedge dA^J.
\end{equation} 
where $K_{IJ}$ is a symmetric integral valued matrix. 
The above most general Abelian Chern-Simons theory includes a particular case: 
\begin{equation}
 S[A,B]=\int_{{M^{3}}} \sum_{I} \frac{N_I}{2\pi}\,B^I\wedge dA^I+
\sum_{I,J} \frac{p_{IJ}}{4\pi}A^I\wedge dA^J
\label{BdAAdA-action}
\end{equation} 
where $p_{IJ}$ is a symmetric integral valued matrix. 
When $p_{IJ}$ is an odd integer, we have the Abelian spin-Chern-Simons theory (considered in detail in \cite{BelovMoore2005ze}).
When $p_{IJ}$ is an even integer, we have the Abelian Chern-Simons theory that are within the 
cohomology group $H^3(\Z_{N_I} \times \Z_{N_J} ,U(1))=\mZ_{N_I} \times \mZ_{N_J} \times \mZ_{N_{IJ}}$ for the Dijkgraaf-Witten theory \cite{1405.7689},
$p_{II} \in \mZ_{N_I}$, $p_{JJ} \in \mZ_{N_J}$ and $p_{IJ} \in \mZ_{N_{IJ}}$.
Here we denote $\mZ_{N_{IJ \dots}} \equiv  \mZ_{\gcd(N_{I}, N_{J},  \dots)}$.

Note that when $K_{II}$ is odd for some $I$, the theory becomes fermionic spin-TQFT that depends on the choice of spin structure. 
A generic collection of line operators supported on $s$ closed disjoint curves $\gamma_I$ embedded in $S^3$ can be realized as follows:
\begin{equation}
 W_{q}[\{\gamma_I\}_{I=1}^s]=
\exp ( i \sum_{I=1}^s e_I\int_{\gamma_I} A_I)\equiv
\exp (i \sum_{I=1}^s e_I\int_{{M^{3}}} A_I\wedge \delta^\perp(\gamma_I) )
\end{equation} 
for some integer numbers $e_I$. As we will see the result, up to a $\pm 1$ sign, only depends on the class of $s$-vector $e$ in the cokernel of the level matrix $K$, that is effectively $ e \in \Z^s/K\Z^s$. Suppose ${M^{3}}=S^3$. The expectation value of $W_{e}[\{\gamma_I\}]$ is then given by a Gaussian integral which \cred{localizes} on the following equations of motion:
\begin{equation}
 \sum_{J}K_{IJ}dA_J=-2\pi \delta^\perp(\gamma_I)
\end{equation} 
which, up to a gauge transformation, can be solved as follows:
\begin{equation}
 A_I=-2\pi \sum_{J}(K^{-1})_{IJ} e_J\delta^\perp(\Sigma_J)
\end{equation} 
where $\Sigma_J$ is a Seifert surface bounded by $\gamma_J$ and we used that  $d\delta^\perp(\Sigma_J)=\delta^\perp(\partial\Sigma)$. Plugging the solution back into the integrand gives us
\begin{multline} \label{Ab-CS-linking}
 \langle W_{e}[\{\gamma_I\}] \rangle
=\exp\left\{ - \pi  i \sum_{I,J} (K^{-1})_{IJ}e_I e_J\int_{S^3}\delta^\perp(\Sigma_I) \wedge d\delta^\perp(\Sigma_J)
\right\}=\\
\exp\left\{ - \pi  i \sum_{I,J} (K^{-1})_{IJ} e_I e_J\text{Lk}(\gamma_I,\gamma_J)
\right\}
\end{multline}
where $\text{Lk}(\gamma_I,\gamma_J)$ is the linking number between $\gamma_I$ and $\gamma_J$, which is by definition equal to the intersection number \cred{$\#(\Sigma_I\cap \gamma_J)$}. 
The physics literature on this invariant dates back to \cite{PhysRevLett.51.2250,Polyakov:1988md}.


\section{$\int BdA+A^3$ in 2+1D, non-Abelian anyons and Milnor's triple linking number}
\label{sec:aaa-theory}

In the 2+1D spacetime, we can consider the following action on a 3-manifold ${M^{3}}$:
\begin{equation}
	S[A,B]=\int_{{M^{3}}}\,\sum_{I=1}^3\frac{N_I}{2\pi}B^I\wedge dA^I+\frac{\bar p}{(2\pi)^2} \,A^1\wedge A^2 \wedge A^3
\end{equation}
where $A^I$ and $B^I$ are 1-form fields. 
Here ${\bar p} \equiv { \frac{N_1 N_2 N_3\;
p_{}}{ N_{123}}}$ with $p \in \mZ_{N_{IJK}}$.   
We have the TQFT that are within the
class $p \in \mZ_{N_{IJK}}$ in the  
cohomology group $H^3(\Z_{N_I} \times \Z_{N_J} \times \Z_{N_K} ,U(1))$ for the Dijkgraaf-Witten theory \cite{1405.7689}.

\begin{figure}[h]
\centering
\includegraphics[scale=1.8]{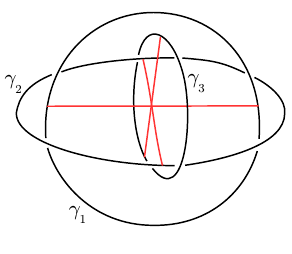}
\caption{Particular choice of surfaces $\Sigma_I$ for Borromean rings. The red lines show pairwise intersections $\Sigma_J\cap \Sigma_K$. The endpoints of the redlines are intersection points which pairs are counted in (\ref{int-pair-count}).}
\label{fig:borromean-rings-int}
\end{figure}

The gauge transformation is:
\begin{equation}
	\begin{array}{c}
		A^I\rightarrow A^I+dg^I \\
		B^I\rightarrow B^I+d\eta^I+\frac{\bar p}{2\pi N_I}\,\epsilon_{IJK}\,\left(A^Jg^K-\frac{1}{2} g^Jdg^K\right).
	\end{array}
\end{equation}
Consider the following observable:
\begin{equation}
	W_{r,q}[\gamma_1,\gamma_2,\gamma_3]=
	\exp\left\{
		i\sum_{I=1}^3\oint_{\gamma_I} q_I\left(B^I+\frac{\bar p}{4\pi\,N_I}\,\epsilon_{IJK}A^J(d^{-1}A^K)\right)+\sum_{J} e_{IJ}A^J
	\right\}
	\label{triple-line-op}
\end{equation}
Where $\gamma_I$ are three pairwise unlinked (and with trivial framing) connected components of a link. The functions $(d^{-1}A^K)$ are defined on link components as follows:
\begin{equation}
	(d^{-1}A^K)(x) \equiv  {\phi^K(x)} \equiv \int\limits_{[x_0,x]_{\gamma_I}}A^K,\qquad x\in\gamma^I
	\label{dinvA}
\end{equation}
where $x_0\in \gamma_I$ is a reference point and $[x_0,x]_{\gamma_I}\subset \gamma_I$ denotes a segment of $\gamma_I$. Note that $\phi^{K}(x)$ is a well defined continuous function on $\gamma_I$ only if $\int_{\gamma_I}A^K=0$, that is the flux of $A^K$ gauge field through $\gamma_I$ vanishes. We assume that this is the case. If such condition is not satisfied $W_{r,q}[\gamma_1,\gamma_2,\gamma_3]$ should be zero instead \cite{He1608.05393}. Later we will generalize this to the case when charges $q_I$, similarly to $e_{IJ}$, form a \cred{general matrix}. 
We are interested in calculating its vacuum expectation value (vev), that is:
\begin{equation}
	\langle W_{q,e}[\gamma_1,\gamma_2,\gamma_3]\rangle =
	\frac{\int \CD A\CD B \;e^{iS[A,B]} W_{q,e}[\gamma_1,\gamma_2,\gamma_3]}
	{\int \CD A\CD B \;e^{iS[A,B]} }.
	\label{vev1}
\end{equation}
As before, $\delta^\perp(\gamma)$ denotes the form distribution supported on $\gamma$ such that $\int_{{M^{3}}} \omega \wedge \delta^\perp(\gamma)=\int_\gamma \omega$ for any $\omega$. Then we can write
\bea
	W_{q,e}[\gamma_1,\gamma_2,\gamma_3] &=&
	\exp\left\{
		i\int_{{M^{3}}} \sum_{I=1}^3\delta^\perp(\gamma_I)\wedge\left[ q_I\left(B^I+\frac{\bar p}{4\pi\,N_I}\sum_{J,K}\epsilon_{IJK}A^J(d^{-1}A^K)\right)+\sum_{J}e_{IJ}A^J\right]
	\right\} \nonumber \\
	&\equiv&
	\exp\left\{
		i\int_{{M^{3}}} \sum_{I=1}^3\delta^\perp(\gamma_I)\wedge\left[ q_I\left(B^I+\frac{\bar p}{4\pi\,N_I}\sum_{J,K}\epsilon_{IJK} d \phi^J\phi^K\right)+\sum_{J}e_{IJ}A^J\right]
	\right\}.\;\;\;\;\;\;\;\;
\eea
Then integrating out $B^I$ in the path integral (\ref{vev1}) imposes the following conditions on $A^I$:
\begin{equation}
	dA^{I}=-\frac{2\pi q_I}{N_I}\,\delta^\perp(\gamma_I)
\end{equation} 
On ${M^{3}}=S^3$ it can be always solved as follows (uniquely modulo the gauge group):
\begin{equation}
	A^{I}=-\frac{2\pi q_I}{N_I}\,\delta^\perp(\Sigma_I)
\end{equation}
where $\Sigma_I$ is a surfcase bounded by $\gamma_I$ (i.e. $\partial\Sigma_I=\gamma_I$). Consider then the value of different terms in the effective action that we obtained after integrating $B^I$ out:
\begin{multline}
	\sum_{I,J}e_{IJ}\int \delta^\perp(\gamma_I) \wedge A_J=
	\sum_{I,J}\frac{2\pi \,e_{IJ} q_J}{N_J}\int \delta^\perp(\gamma_I) \wedge \delta^\perp(\Sigma_J)=\\
	=\sum_{I,J}\frac{2\pi \,e_{IJ} q_J}{N_J}\, \# (\gamma_I \cap \Sigma_J)\equiv
	\sum_{I,J}\frac{2\pi \,e_{IJ} q_J}{N_J}\, \text{Lk}(\gamma_I, \gamma_J).
\end{multline}
The assumption that there is no flux of $A^I$ gauge field through any $\gamma_J$ for any pair $I,J$ implies that in order to get a non-vanishing expectation value all pairwise linking numbers should be zero: $\text{Lk}(\gamma_I, \gamma_J)=0$. 
\begin{multline}
	\int \frac{\bar p}{(2\pi)^2} \,A^1\wedge A^2 \wedge A^3=-\frac{2\pi\,\bar p\, q_1 q_2 q_3}{N_1N_2N_3}\int \delta^\perp(\Sigma_1) \wedge \delta^\perp(\Sigma_2) \wedge \delta^\perp(\Sigma_3)=\\
	=-\frac{2\pi\,{\bar p}\, q_1 q_2 q_3}{N_1N_2N_3}\,\# (\Sigma_1 \cap \Sigma_2 \cap \Sigma_3)
\end{multline}
where intersection numbers are, as usual, counted with signs determined by orientation. Denote $(-1)^{\epsilon(a)}$ the sign corresponding to the orientation of the intersection at point $a$. Consider (\ref{dinvA}):
\begin{equation}
	-\frac{N_K}{2\pi q_K}\,(\phi^K)(x)=\int\limits_{x_0\rightarrow x\text{ along }\gamma_I}\delta^\perp(\Sigma_K)=\# ([x_0,x]_{\gamma_I} \cap \Sigma_K)\equiv 
	\sum_{a\in ([x_0,x]_{\gamma_I} \cap \Sigma_K)}(-1)^{\epsilon(a)}
\end{equation}
which is unambiguously defined because $\text{Lk}(\gamma_I,\gamma_K)=0$.
Then
\begin{multline}
	\sum_{I,J,K}\frac{{\bar p}\, q_I\,\epsilon_{IJK}}{4\pi\,N_I}\int \delta^\perp(\gamma_I)\wedge A^J(d^{-1}A^K)=
	\sum_{I,J,K}\frac{{\bar p}\, q_I q_J\,\epsilon_{IJK}}{2N_IN_J}\int \delta^\perp(\gamma_I)\wedge \delta^\perp(\Sigma_J)\,(d^{-1}A^K)=\\
	\sum_{I,J,K}\frac{\pi\, {\bar p} \,q_I q_J q_K\,\epsilon_{IJK}}{N_IN_JN_K}\sum_{b\;\in\; \gamma_I \cap \Sigma_K}(-1)^{\epsilon(b)}\sum_{a\in ([x_0,x_b]_{\gamma_I} \cap \Sigma_K)}(-1)^{\epsilon(a)}=\\
	\frac{\pi\, {\bar p} \, q_1 q_2 q_3}{N_1N_2N_3}\sum_{I,J,K}\epsilon_{IJK}
	\sum_{\scriptsize\begin{array}{c}a\in \gamma_{I}\cap \Sigma_K \\ b\in \gamma_{I}\cap \Sigma_J\\
x_b>x_a \end{array}}(-1)^{\epsilon(a)+\epsilon(b)},
\label{int-pair-count}
\end{multline}
where the ordering of intersection points $a,b\,\in\, \gamma_I$ (that is the condition $x_b>x_a$) is done relative to the previously chosen reference point $x_0\in \gamma_I$. Finally we have:
\begin{equation}
	\langle W_{q,e}[\gamma_1,\gamma_2,\gamma_3]\rangle
	=
	\exp\left\{
	-\frac{2\pi i\,{\bar p}\, q_1 q_2 q_3}{N_1N_2N_3}\,\bar{\mu}(\gamma_1,\gamma_2,\gamma_3)
	\right\}
	=
	\exp\left\{
	-\frac{2\pi i\,p\,q_1q_2 q_3}{N_{123}}\,\bar{\mu}(\gamma_1,\gamma_2,\gamma_3)
	\right\}
	\label{triple-line-vev-result}
\end{equation}
where
\begin{equation}
	\bar{\mu}(\gamma_1,\gamma_2,\gamma_3)=
		\# (\Sigma_1 \cap \Sigma_2 \cap \Sigma_3)-\frac{1}{2}\sum_{I,J,K}\epsilon_{IJK}
		\sum_{\scriptsize\begin{array}{c}a\in \gamma_{I}\cap \Sigma_K \\ b\in \gamma_{I}\cap \Sigma_J\\
	x_a>x_b \end{array}}(-1)^{\epsilon(a)+\epsilon(b)}
		\label{triple-linking}
\end{equation}
is exactly the geometric formula for Milnor's $\bar{\mu}$ invariant or Milnor's triple linking number \cite{geom-triple-linking}. It is easy to evaluate for the Borromean rings link. Consider realization of Borromean rings shown in Figure \ref{fig:borromean-rings-int} with natural choice of Seifert surfaces $\Sigma_I$ lying in three pairwise orthogonal planes. It is easy to see that the first term in (\ref{triple-linking}) is $1$ while all other terms vanish. That is
\begin{equation}
	\bar{\mu}(\gamma_1,\gamma_2,\gamma_3)=1.
\end{equation}
As an example, in the corresponding link figure shown in Table \ref{table:TQFTlink}, we mean the braiding process of three particle excitations described in 
\cite{CWangMLevin1412.1781,1602.05951,RyuTiwariChen1603.08429}.

When the coefficients $q$ in (\ref{triple-line-op}) form a general 
matrix $q_{IJ}$ (similarly to the coefficients $e_{IJ}$) we have the $(\det q)$ instead of $q_1 q_2 q_3$ in (\ref{triple-line-vev-result})\footnote{Assuming trivial framing of link components}. Lastly, we remark that this 2+1D theory with a cubic interacting action can host 
non-Abelian anyons\cite{deWildPropitius:1995cf,Wang1404.7854,CWangMLevin1412.1781, He1608.05393} with non-Abelian statistics. 
\cred{
The attempt to derive the Milnor's $\bar{\mu}$ invariant from Chern-Simons\cgreen{-like} field theory dates back to \cite{Ferrari0210100, Leal0704.2429} and recently summarized in \cite{Franco1411.6429}.\footnote{We thank Franco Ferrari for bringing us attention to the earlier work on $\int BdA+A^3$ theory.}
However, our approach is rather different and is generally based on Poincar\'e duality and the intersection theory.}
We note that the theory of
$\int\,\sum_{I=1}^3\frac{N_I}{2\pi}B^I\wedge dA^I+\frac{\bar p}{(2\pi)^2} \,A^1\wedge A^2 \wedge A^3$ is equivalent to the non-Abelian discrete gauge theory of the dihedral group $D_4$ (with the $D_4$ group of order 8) \cite{deWildPropitius:1995cf, Wang1404.7854}. 

\section{$\int BdA+A^2dA$ in 3+1D and the triple linking number of 2-surfaces}
\label{sec:AAdA}

In the 3+1D spacetime, consider the following action on a 4-manifold ${M^{4}}$:
\begin{equation}
	S[A,B]=\int_{{M^{4}}}\,\sum_{I=1}^3\frac{N_I}{2\pi}B^I\wedge dA^I+\frac{{\bar p}}{(2\pi)^2} \,A^1\wedge A^2 \wedge dA^3
\end{equation}
where $A^I$ and $B^I$ are 1- and 2-form gauge fields respectively. 
Here ${\bar p} \equiv { \frac{N_1 N_2 \;
p_{}}{ N_{12}}}$ with $p \in \mZ_{N_{123}}$.   
We have the TQFT that are within the
class $p \in \mZ_{N_{123}}$ in the  
cohomology group $H^4(\Z_{N_1} \times \Z_{N_2} \times \Z_{N_3} ,U(1))$ for the Dijkgraaf-Witten theory \cite{1405.7689}.

Let us introduce an antisymmetric matrix $\epsilon^{IJ}$ such that $\epsilon^{12}=-\epsilon^{21}=1$ and all other elements are zero. The gauge transformation then reads:
\begin{equation}
	\begin{array}{c}
		A^I\rightarrow A^I+dg^I, \\
		B^I\rightarrow B^I+d\eta^I+\frac{{\bar p}}{2\pi N_I}\,\epsilon^{IJ}dg^J\wedge A^3. 
	\end{array}
\end{equation}
Consider the following gauge invariant observable:
\begin{equation}
	W_{q}[\Sigma_1,\Sigma_2,\Sigma_3]=
	\exp(
		i\sum_{I=1}^3 q_I\left\{\int_{\Sigma_I} B^I+ \sum_{J} \frac{{\bar p}\,\epsilon^{IJ}}{2\pi\,N_I}\int_{\CV_I} A^J\wedge dA^3\right\}).
	\label{triple-surface-op}
\end{equation}
Where $\Sigma_I$ are three non-intersecting surfaces in ${M^{4}}$ and $\CV_I$ are some 3D submanifolds that are bounded by them, that is $\partial\CV_I=\Sigma_I$. Such $\CV_I$ are usually called Seifert hyper-surfaces. As before, we are interested in calculating its vev, that is:
\begin{equation}
	\langle W_{q}[\Sigma_1,\Sigma_2,\Sigma_3]\rangle =
	\frac{\int \CD A\CD B \;e^{iS[A,B]} W_{q}[\Sigma_1,\Sigma_2,\Sigma_3]}
	{\int \CD A\CD B \;e^{iS[A,B]} }.
	\label{vev1}
\end{equation}
Using $\delta$-forms we can write it as follows
\begin{equation}
	 W_{q}[\Sigma_1,\Sigma_2,\Sigma_3]=
	\exp (i\sum_{I=1}^3 q_I\int_{{M^{4}}}\left\{\delta^\perp(\Sigma_I)\wedge B^I+\frac{{\bar p}\,\epsilon^{IJ}}{2\pi\,N_I}\,\delta^\perp(\CV_I)\wedge A^J\wedge dA^3\right\})
\end{equation}
Then integrating out $B^I$ in the path integral (\ref{vev1}) imposes the following conditions on $A^I$:
\begin{equation}
	dA^{I}=-\frac{2\pi q_I}{N_I}\,\delta^\perp(\Sigma_I)
\end{equation} 
On ${M^{4}}=S^4$ it can be always solved as follows (uniquely modulo the gauge group):
\begin{equation}
	A^{I}=-\frac{2\pi q_I}{N_I}\,\delta^\perp(\CV_I')
\end{equation}
where $\CV_I'$ is any 3D \cred{hypersurface} bounded by $\Sigma_I$. Without loss of generality we can choose $\CV_I'=\CV_I$. Consider the value of different terms in the effective action that we obtained after integrating $B^I$ out:
\begin{multline}
	\int \frac{{\bar p}}{(2\pi)^2} \,A^1\wedge A^2 \wedge dA^3=-\frac{2\pi\,{\bar p}\, q_1 q_2 q_3}{N_1N_2N_3}\int \delta^\perp(\CV_1) \wedge \delta^\perp(\CV_2) \wedge \delta^\perp(\Sigma_3)=\\
	=-\frac{2\pi\, {\bar p} \,q_1 q_2 q_3}{N_1N_2N_3}\,\# (\CV_1 \cap \CV_2 \cap \Sigma_3)
\end{multline}
\begin{multline}
	\sum_{I=1}^3 q_I\int_{{M^{4}}}\frac{{\bar p}\,\epsilon^{IJ}}{2\pi\,N_I}\,\delta^\perp(\CV_I)\wedge A^J\wedge dA^3=
	\sum_{I=1}^3\frac{2\pi {\bar p}\,\epsilon^{IJ} q_I q_J q_3}{N_IN_JN_3}\int_{{M^{4}}}\,\delta^\perp(\CV_I)\wedge \delta^\perp(\CV_J)\wedge \delta^\perp(\Sigma_3)=\\
	=\frac{4\pi {\bar p}\, q_1 q_2 q_3}{N_1N_2N_3}
	\,\# (\CV_1 \cap \CV_2 \cap \Sigma_3).
\end{multline}

 Finally we have:
\begin{equation}
	\langle W_{q}[\Sigma_1,\Sigma_2,\Sigma_3]\rangle
	=
	\exp\left\{
	\frac{2\pi i\, {\bar p} \, q_1 q_2 q_3}{N_1N_2N_3}\,\text{Tlk}(\Sigma_1,\Sigma_3,\Sigma_2)
	\right\}
	=
	\exp\left\{
	\frac{2\pi i\,p\,q_1 q_2 q_3}{N_{123}}\,\text{Tlk}(\Sigma_1,\Sigma_3,\Sigma_2)
	\right\}
	\label{triple-surface-vev-result}
\end{equation}
where
\begin{equation}
	\text{Tlk}(\Sigma_1,\Sigma_3,\Sigma_2)\equiv \# (\CV_1 \cap \CV_2 \cap \Sigma_3)
		\label{triple-surface-linking}
\end{equation}
is the triple linking number of surfaces (defenition (4) in \cite{surface-triple-linking}).

\begin{figure}[!t]
\centering
\includegraphics[scale=1.2]{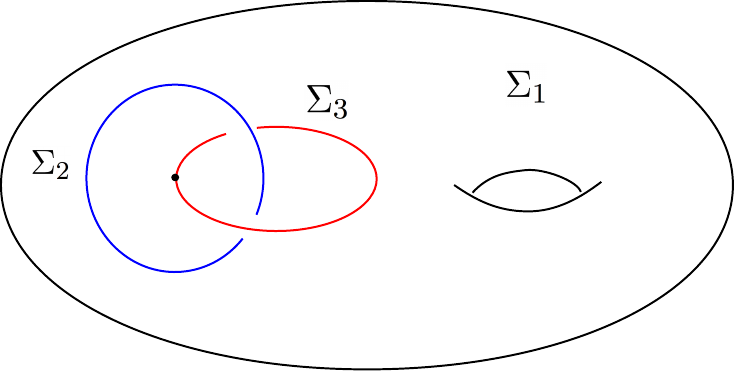}
\caption{\cred{An example of configuration with a triple linking number Eq.(\ref{triple-surface-linking}) of three 2-surfaces being $\text{Tlk}(\Sigma_1,\Sigma_3,\Sigma_2)\equiv \# (\CV_1 \cap \CV_2 \cap \Sigma_3)=1$. (The same link figure is shown in Table \ref{table:TQFTlink}.)
Take the $\Sigma_2,\Sigma_3$ to be a spun of Hopf link with Seifert hypersurfaces $\CV_2,\CV_3$ being \cgreen{spuns} of Seifert surfaces. The surface $\Sigma_1$ is a torus embedded at a fixed value of the spin angle and encircling the Hopf link. Choose Seifert hypersurface $\CV_1$ to be the interior the torus $\Sigma_1$. 
The intersection $\# (\CV_1 \cap \CV_2 \cap \Sigma_3)=1$ contains one point shown bold in the figure.}
}
\label{fig:Hopf-spun-int}
\end{figure}

Similarly one can consider theory
\begin{equation}
	S[A,B]=\int_{{M^{4}}}\,\sum_{I=1}^2\frac{N_I}{2\pi}B^I\wedge dA^I+\frac{{\bar p}}{(2\pi)^2} \,A^1\wedge A^2 \wedge dA^2
\end{equation}
which represents a non-trivial element of \cred{$H^4(\Z_{N_1} \times \Z_{N_2}  ,U(1))$}. The analogous operator supported in a triple of surfaces $\Sigma_{1},\Sigma_{2},\Sigma_{2}'$ reads:
\begin{equation}
	W_{q}[\Sigma_1,\Sigma_2,\Sigma_2']=
	\exp(
		iq_1\int_{\Sigma_1} B^1+iq_2\int_{\Sigma_2}B_2+iq_2'\int_{\Sigma_2'}B_2 +\ldots)
\end{equation}
where dots denote appropriate gauge invariant \cred{completions} similar to the ones in (\ref{triple-surface-op}). The resulting expectation value is as follows
\begin{equation}
	\langle W_{e,q}[\Sigma_1,\Sigma_2,\Sigma_2']\rangle
	=
	\exp\left\{
	\frac{2\pi i\, {\bar p} \, q_1 q_2 q_3}{N_1N_2N_3}\,[\text{Tlk}(\Sigma_1,\Sigma_2,\Sigma_2')+\text{Tlk}(\Sigma_1,\Sigma_2',\Sigma_2)]
	\right\}
\end{equation}
assuming $\Sigma_2,\Sigma_2'$ have trivial framing. 

\cred{As an example, in the corresponding link figure shown in Table \ref{table:TQFTlink} as well \cgreen{in} Fig. \ref{fig:Hopf-spun-int}, we mean the braiding process of three string excitations described in 
\cite{Wang1403.7437,Jiang:2014ksa,Wang1404.7854,1602.05951,RyuTiwariChen1603.08429}.
In this configuration \cgreen{shown in} Fig. \ref{fig:Hopf-spun-int}, we have $\text{Tlk}(\Sigma_1,\Sigma_3,\Sigma_2)=1$, $\text{Tlk}(\Sigma_2,\Sigma_3,\Sigma_1)=-1$, 
$\text{Tlk}(\Sigma_3,\Sigma_2,\Sigma_1)=-1$, $\text{Tlk}(\Sigma_1,\Sigma_2,\Sigma_3)=1$, and finally 
$\text{Tlk}(\Sigma_2,\Sigma_1,\Sigma_3)=\text{Tlk}(\Sigma_3,\Sigma_1,\Sigma_2)=0$.
The TQFTs with $A_1A_2dA_3$ \cgreen{term} can detect this link, and 
also $A_2A_1dA_3,\,A_3A_1dA_2,\, A_1A_3dA_2$ can detect this link, but neither $A_2A_3dA_1$ nor $A_3A_2dA_1$ can detect this link configuration.
}

\cred{Reader can find Ref.\cite{RyuTiwariChen1603.08429} for a related study for this theory.
We also note that Ref.\cite{Bi:2014vaa} applies non-linear sigma model \cgreen{descriptions as an alternative,} without using $\int BdA+A^2dA$ theory to study the 3 string braiding process, 
limited \cgreen{to} a more restricted case $N_1= N_2=N_3= 2$. 
Here we had considered more generic \cgreen{levels}.}

\section{$\int BdA+A^4$ in 3+1D, non-Abelian strings and 
the quadruple linking number of 2-surfaces}
\label{sec:A4-theory}

In the 3+1D spacetime, we can also consider the following action on a 4-manifold ${M^{4}}$:
\begin{equation}
	S[A,B]=\int_{{M^{4}}}\,\sum_{I=1}^4\frac{N_I}{2\pi}B^I\wedge dA^I+\frac{\bar p}{(2\pi)^3} \,A^1\wedge A^2 \wedge A^3 \wedge A^4
\end{equation}
where $A^I$ and $B^I$ are 1- and 2-form gauge fields respectively. 
Here ${\bar p} \equiv { \frac{N_1 N_2 N_3 N_4\;
p_{}}{ N_{1234}}} $ with $p \in \mZ_{N_{1234}}$.   
We have the TQFT that are within the
class $p \in \mZ_{N_{1234}}$ in the  
cohomology group $H^4(\Z_{N_1} \times \Z_{N_2} \times \Z_{N_3} \times \Z_{N_4} ,U(1))$ for the Dijkgraaf-Witten theory \cite{1405.7689}.

The gauge transformation reads (see the exact transformation to all order in \cite{1602.05951}):
\begin{equation}
	\begin{array}{c}
		A^I\rightarrow A^I+dg^I \\
		B^I\rightarrow B^I+d\eta^I-\sum_{J,K,L}\frac{{\bar p}}{2(2\pi)^2N_I}\,\epsilon^{IJKL}A^J\,A^K\,g^L
	\end{array}
\end{equation}
where $\epsilon^{IJKL}$ is an absolutely anti-symmetric tensor. Consider the 
 surface operators supported on 4 different non-intersecting surfaces $\Sigma_I,\, I=1,\ldots,4$ which we \textit{formally} write as follows:
\begin{equation}
	W_{q}[\Sigma_1,\Sigma_2,\Sigma_3,\Sigma_4]=
	\exp
		(i\sum_{I=1}^4 q_I\left\{\int_{\Sigma_I} B^I+\sum_{J,K,L}\frac{{\bar p}\,\epsilon^{IJKL}}{3!\,(2\pi)^2\,N_I}A^I A^J d^{-1}A^L\right\})
	\label{quadruple-surface-op}
\end{equation}
To be more specific, consider the surface operator supported on $\Sigma_1$:
\begin{equation}
	\exp
		(i q_1\left\{\int_{\Sigma_1} B^1+\sum_{J,K,L\neq 1}\frac{{\bar p}\,\epsilon^{1JKL}}{3!\,(2\pi)^2\,N_I}A^I A^J d^{-1}A^L\right\}).
\end{equation}
What we mean by this expression is the following. If in the path integral we first integrate out $B^2,B^3,B^4$ (which do not appear in the surface operator supported on $\Sigma_1$), this imposes conditions 
\begin{equation}
	dA^J=0,\,J=2,3,4.\label{flat-cond}
\end{equation}
 If $\Sigma_1$ has a non-zero genus as a Riemann surface, we can always represent it by a polygon $\tilde\Sigma_1$ 
 (which is topologically a disk) with appropriately glued boundary. 
 Choose a point $x_*^{(1)}\in \tilde\Sigma_1$ and define $d^{-1}A^I|_{\Sigma_1}\equiv \phi^I(x)\equiv\int_{x_*^{(1)}}^x A^I,\,x\in \Sigma_1$ where the integral is taken along a path in $\tilde\Sigma_1$. 
 It does not depend on the choice of the path in $\tilde\Sigma_1$ 
 due to (\ref{flat-cond}). The choice of simply connected $\tilde\Sigma_1$ 
 representing $\Sigma_1$ is similar to the global choice of the path in $\gamma_I$ for line operators in section \ref{sec:aaa-theory}. The surface operator that can be expressed as
\begin{equation}
	\exp (
		i q_1\left\{\int_{\Sigma_1} B^1+\sum_{J,K,L\neq 1}\frac{{\bar p}\,\epsilon^{1JKL}}{3!\,(2\pi)^2\,N_I}d\phi^I d\phi^J \phi^L\right\}).
\end{equation}
It is easy to see that it is invariant under the gauge transformations
\begin{equation}
	\begin{array}{c}
	\phi^I\rightarrow \phi^I+g^I \\
		B^1\rightarrow B^1+d\eta^1-\sum_{J,K,L\neq 1}\frac{{\bar p}}{2(2\pi)^2N_1}\,\epsilon^{1JKL}d\phi^J\,d\phi^K\,g^L
	\end{array}
\end{equation}
up to boundary terms supported on $\partial \tilde\Sigma_1$. 
The presence of such terms and the dependence on the choice of $\tilde\Sigma_1$ 
in general makes such surface operator ill defined. However for field configurations with certain restriction the boundary terms vanish and the ambuguity goes away. This is similar to the situation in Sec. $\ref{sec:aaa-theory}$, where $d^{-1}A_K$ is a well defined continuous function on $\gamma_I$ only if the pairwise linking numbers vanish. In particular, we will need to require that all triple linking numbers $\text{Tlk}(\Sigma_I,\Sigma_J,\Sigma_K)$ are zero. If the ambiguity is present, the operator should vanish instead, similarly to the case considered in Sec. \ref{sec:aaa-theory}.  In the examples below there will be no such ambiguity.

\begin{figure}[!h]
\centering
\includegraphics[scale=1.4]{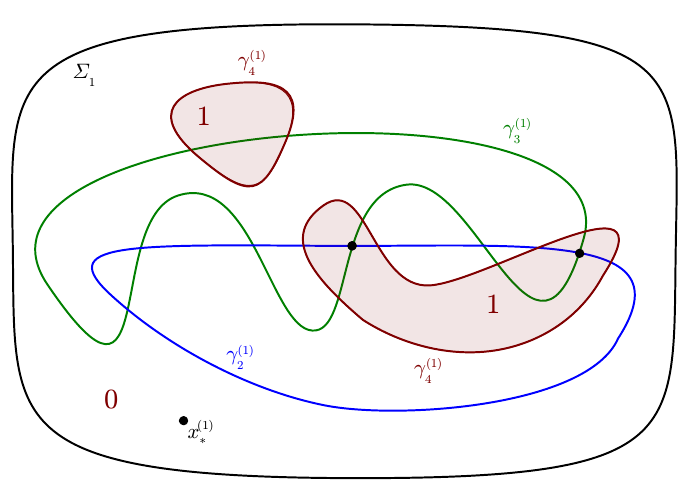}
\caption{An example of computing $t(\Sigma_1;\CV_2,\CV_3,\CV_4)$ by formula (\ref{t-surface-formula}). The red numbers $0,1$ denote the value of the weight with which the intersection points of $\gamma_2^{(1)}$ with $\gamma_3^{(1)}$ are counted in different domains separated by $\gamma_4^{(1)}$. The points $a\in \gamma^{(1)}_2\cap \gamma^{(1)}_3$ that enter into the sum with non-zero weight are shown as bold black points.  }
\label{fig:qlinking-t-counting}
\end{figure}

As before, let $\CV_I$ be Seifert hypersurfaces such that $\partial\CV_I=\Sigma_I$. Then the integrating out all $B^I$ implies:
\begin{equation}
	dA^{I}=-\frac{2\pi q_I}{N_I}\,\delta^\perp(\Sigma_I),\qquad A^{I}=-\frac{2\pi q_I}{N_I}\,\delta^\perp(\CV_I).
\end{equation} 
The effective action is then given by the quadruple intersection number of the Seifert hypersurfaces
\begin{multline}
	\int \frac{ {\bar p}}{(2\pi)^2} \,A^1\wedge A^2 \wedge A^3\wedge A^4=\frac{2\pi\,{\bar p}\,q_1 q_2 q_3 q_4}{N_1N_2N_3N_4}\int \delta^\perp(\CV_1) \wedge \delta^\perp(\CV_2) \wedge \delta^\perp(\CV_3)\wedge \delta^\perp(\CV_4)=\\
	=\frac{2\pi\,{\bar p} \, q_1 q_2 q_3 q_4}{N_1N_2N_3N_4}\,\# (\CV_1 \cap \CV_2 \cap \CV_3\cap \CV_4)
\end{multline}
The contribution of the surface operator supported on $\Sigma_I$ reads
\begin{multline}
	-q_I\sum_{J,K,L}\int_{{M^{4}}}\frac{{\bar p} \; q_J \; q_K \; q_L\,\epsilon^{IJKL}}{3!\,(2\pi)^2\,N_IN_JN_KN_L}\,\delta^\perp(\Sigma_I)\wedge\delta^\perp(\CV_J)\wedge \delta^\perp(\CV_K)\; \left.d^{-1}\delta^\perp(\CV_L)\right|_{\Sigma_I}=\\
	-\frac{2\pi\,{\bar p} \, q_1 q_2 q_3 q_4}{N_1N_2N_3N_4}
\sum_{J,K,L\neq 4} \frac{t(\Sigma_I;\CV_J,\CV_K,\CV_L)}{3!}
\end{multline}
where $t(\Sigma_I;\CV_J,\CV_K,\CV_L)$ is defined as follows. Consider  $\gamma^{(I)}_{J}\equiv (\Sigma_I\cap \CV_J)$, oriented, not necesserily connected, curves on surface $\Sigma_I$. Then\footnote{Again, we assume that configuration of surfaces is such that there is no ambiguity in such expression, i.e. no dependence on the choice of $\tilde{\Sigma}_1$ and $x_*^{(1)}\in \tilde{\Sigma}_1$. Otherwise the result should be zero: $\langle W \rangle$.}
\bea
 t(\Sigma_1;\CV_2,\CV_3,\CV_4)\equiv
\sum_{a\,\in \,\gamma^{(1)}_2\cap \gamma^{(1)}_3 }(-1)^{\epsilon(a)} 
\,\left[\#\text{ of times one needs to cross $\gamma^{(1)}_4$ to reach $a$ from $x_*^{(1)}$}\right],
\label{t-surface-formula}
\eea
where as before, $(-1)^{\epsilon(a)}=\pm 1$ 
depends on the orientation of the intersection of $\gamma^{(1)}_2$ with $\gamma^{(1)}_3$. The crossings with $\gamma^{(1)}_4$ are also counted with signs. See Fig. \ref{fig:qlinking-t-counting}. 
Note that if there is no ambiguity in defining $t(\Sigma_1;\CV_2,\CV_3,\CV_4)$ (i.e. no dependence on the choice of polygon $ \tilde\Sigma_1$
and the reference point $x_*^{(1)}$) it is antisymmetric 
with respect to exchange of $\CV_2,\CV_3,\CV_4$.

\begin{figure}[!h]
\centering
\includegraphics[scale=1.6]{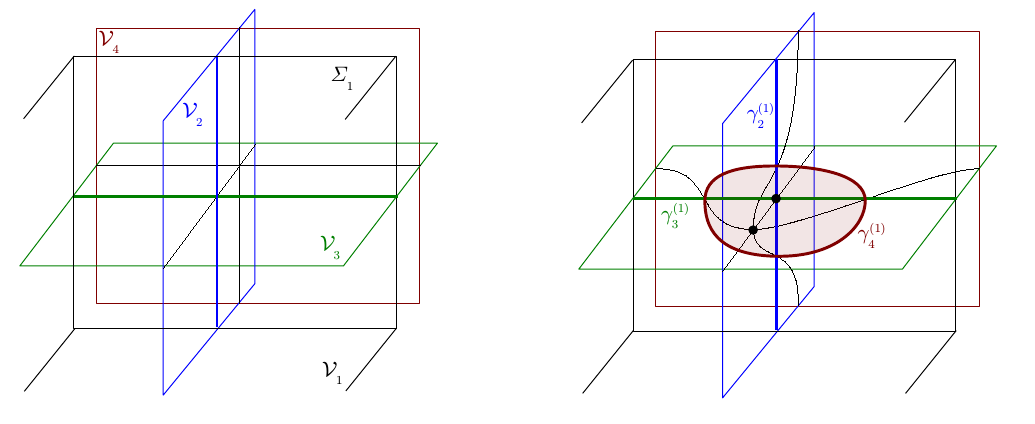}
\caption{An illustration of invariance of (\ref{quadruple-surface-linking}) under deformation of Seifert hypersurfaces $\CV_I$. A local configuration of $\Sigma_1,\CV_2,\CV_3,\CV_4$ in shown in $\R^4\cong \R^3 \times \R_\text{time}$, where $\R_\text{time}$ is not shown in the picture. The hypersurfaces $\CV_2,\CV_3,\CV_4$ are locally repsented by planes $\times \R_\text{time}$, while $\Sigma_1$ is locally a plane $\times$ point and $\CV_1$ is locally a half $\R^3$ bounded by $\Sigma^1$ and spanned in the direction of the reader. The right side shows a local deformation of $\CV_4$ which results in increasing both $\# (\CV_1 \cap \CV_2 \cap \CV_3\cap \CV_4)$ and $t(\Sigma_1;\CV_2,\CV_3,\CV_4)$ by 1 (the contributing intersection points are shown bold and black). The total sum (\ref{quadruple-surface-linking}) stays intact.}
\label{fig:qlinking-qlinking-move}
\end{figure}
\begin{figure}[!h]
\centering
\includegraphics[scale=1.5]{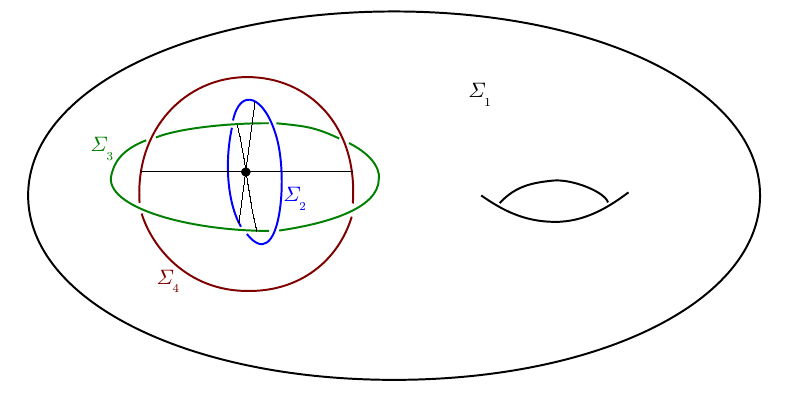}
\caption{An example of configuration with a quadruple linking number Eq.(\ref{quadruple-surface-linking}) being $\text{Qlk}(\Sigma_1,\Sigma_2,\Sigma_3,\Sigma_4)=1$. Take the triple $\Sigma_2,\Sigma_3,\Sigma_4$ to be a spun of Borromean rings with Seifert hypersurfaces $\CV_2,\CV_3,\CV_4$ being the spun of Seifert surfaces in Fig. \ref{fig:borromean-rings-int}. The surface $\Sigma_1$ is a torus embedded at a fixed value of the spin angle and encircling the Borromean rings. Choose Seifert hypersurface $\CV_1$ to be the interior the torus $\Sigma_1$. It is easy to see that for this choice of Seifert hypersurfaces all $t(\Sigma_I;\CV_J,\CV_K,\CV_L)$ vanish just because for each $\Sigma_I$ one of the three curves $\gamma_J^{(I)}\,,J\neq I$ is empty. The quadruple intersection $\# (\CV_1 \cap \CV_2 \cap \CV_3\cap \CV_4)=1$ contains one point shown bold in the figure.
}
\label{fig:borromean-rings-spun-int}
\end{figure}

 Finally we have:
\begin{multline}
	\langle W_{q}[\Sigma_1,\Sigma_2,\Sigma_3,\Sigma_4]\rangle=
	\exp\left\{
	\frac{2\pi i\,{\bar p}\,q_1 q_2 q_3q_4}{N_1N_2N_3N_4}\,\text{Qlk}(\Sigma_1,\Sigma_2,\Sigma_3,\Sigma_4)
	\right\}\\
	=\exp \left\{
	\frac{2\pi i\,p\,q_1 q_2 q_3 q_4}{N_{1234}}\,\text{Qlk}(\Sigma_1,\Sigma_2,\Sigma_3,\Sigma_4)
	\right\}
	\label{quadruple-surface-vev-result}
\end{multline}
where we define the \textit{quadruple linking number} of four 2-surfaces as follows:
\begin{multline}
	\text{Qlk}(\Sigma_1,\Sigma_2,\Sigma_3,\Sigma_4)\equiv 
\# (\CV_1 \cap \CV_2 \cap \CV_3\cap \CV_4)\\
-t(\Sigma_1;\CV_2,\CV_3,\CV_4)
+t(\Sigma_2;\CV_3,\CV_4,\CV_1)
-t(\Sigma_3;\CV_4,\CV_1,\CV_2)
+t(\Sigma_4;\CV_1,\CV_2,\CV_3).
		\label{quadruple-surface-linking}
\end{multline}

It is very similar to the geometric definition Milnor's triple linking number of a 3-component link in $S^3$ considered in Section \ref{sec:aaa-theory}. Each term in the sum is not a topological invariant (that is invariant under ambient isotopy) of embedded quadruple of surfaces $\Sigma_{1,2,3,4}\subset S^4$, since it depends on the choice of Seifert hypersurfaces $\CV_I$. However, their sum is. One can easily check its invariance under basic local deformation moves, see Fig. \ref{fig:qlinking-qlinking-move}. A particular example with quadruple linking number 1 is shown in Fig. \ref{fig:borromean-rings-spun-int}.

Lastly, we remark that this 3+1D theory with a quartic interacting action can host non-Abelian strings\cite{Wang1404.7854, CWangMLevin1412.1781, RyuTiwariChen1603.08429} with non-Abelian statistics. 
As an example, in the corresponding link figure shown in Table \ref{table:TQFTlink} and Fig. \ref{fig:borromean-rings-spun-int}, we mean the braiding process of four string excitations described in 
\cite{CWangMLevin1412.1781,1602.05951,RyuTiwariChen1603.08429}.


\section{$\int BdA+BB$ in 3+1D and the intersection number of open surfaces}
\label{sec:BB-theory}

In the 3+1D spacetime, 
 one can consider the following action on a 4-manifold ${M^{4}}$:
\begin{equation}
	S[A,B]=\int_{{M^{4}}}\,\sum_{I=1}^s\frac{N_I}{2\pi}B^I\wedge dA^I+\sum_{I,J=1}^s\frac{p_{IJ}N_IN_J}{4\pi N_{IJ}}\,B^I\wedge B^J 
	\label{BB-action}
\end{equation}
where $A^I$ and $B^I$ are 1- and 2-form fields respectively and $N_{IJ} \equiv \gcd(N_I,N_J)$. 
We make a choice on the symmetric integral quadratic form $p_{IJ} \in \mZ$.
This TQFT is beyond the Dijkgraaf-Witten group cohomology theory.

The gauge transformation reads:
\begin{equation}
	\begin{array}{c}
		A^I\rightarrow A^I+dg^I -\sum_J \frac{p_{IJ}N_J\eta^J}{N_{IJ}}\\
		B^I\rightarrow B^I+d\eta^I.
	\end{array}
\end{equation}
Note that if the diagonal elements $p_{II}$ \cred{and the integer $N_I$} are odd, $e^{iS}$ is invariant under large gauge transformations only if ${M^{4}}$ has even intersection form. Equivalently, it is a spin 4-manifold.

Consider the following gauge invariant operator supported on closed surfaces $\Omega_I$ and surfaces $\Sigma^I$ with boundaries $\gamma^I$:
\begin{equation}
	W_{e,q}[\{\Sigma^I\},\{\Omega^J\}]=
	\exp\left[
		i\sum_{I}q_I\int_{\Omega_I} B^I+i\sum_{I} e_I
		\left\{
		\int_{\gamma_I} A^I
		+\int_{\Sigma_I} \frac{p_{IJ}N_J}{N_{IJ}}\,B_J
		\right\}
		\right]
	\label{B2-surface-op}
\end{equation}
where $q_I, e_I\in\Z$ (the expectation value, up to a sign, will depend only on their value modulo $N_I$) are integral weights (charges). 
{Since the charge curve ${\gamma_I}$ of $A^I$ must bound the surface ${\Sigma_I}$ of $B_J$,
we learn that the $\int BdA+BB$ theory is a higher-form gauge theory where particles must have strings attached.}

\begin{figure}[h]
\centering
\includegraphics[scale=2]{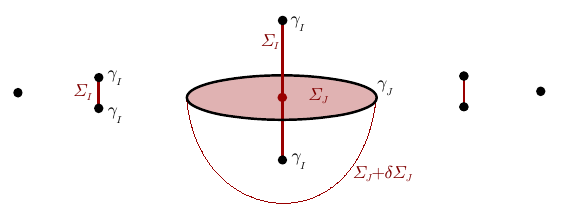}
\caption{An example of a pair of surfaces $\Sigma_I,\Sigma_J\in S^4$ bounded by curves $\gamma_{I},\gamma_J$ with intersection number $\#(\Sigma_I\cap \Sigma_J)=1$. The configuration of line/surface operaors in $S^4$  is represented as a ``movie,'' where the horizontal direction
is the time axis moving from the left to the right. The closed curve $\gamma_I$ is represented by a pair of points created in 3-dimensional space and then annihilated.  Seifert surface $\Sigma_I$ is represented by line in the 3-dimensional space connecting the points in the pair. Seifert surface $\Sigma_J$ is represented by a disc appearing at a fixed moment in ``time''. The red bold point depicts the point contributing to the intersection number $\#(\Sigma_I\cap \Sigma_J)=1$. The shifted surface $\Sigma_J+\delta\Sigma_J$ does not intersect $\Sigma_I$, which is consistent with the fact that the 2D closed surface $\delta\Sigma_J$ has a non-trivial linking number with the 1d closed curve $\gamma_I$: 
\cred{$\text{Lk}(\gamma_I,\delta \Sigma_J)=1$}.
}
\label{fig:BB-int-number}
\end{figure}

\begin{figure}[!h]
\centering
\includegraphics[scale=1.7]{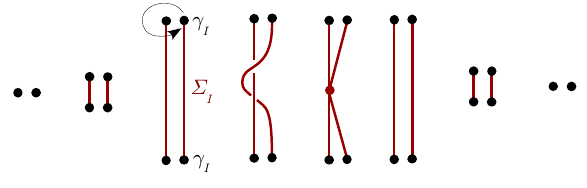}
\caption{An example of a framing choice which results in the self-intersection number $\#(\Sigma_I\cap \Sigma_I)=1$. The configuration of line/surface operaors in $S^4$  is represented as a ``movie''. The closed curve $\gamma^I$ and its sligthly shifted copy are represented by a pairs of points created in 3-dimensional space and then annihilated.  Seifert surfaces are represented by lines in a 3-dimensional space connecting the points in the pairs. The red bold point depicts the point contributing to the self-intersection number $\#(\Sigma_I\cap \Sigma_I)=1$. Note that when \cgreen{both $p_{II}$ and $N_I$ are odd,} such configuration result in \cred{$\langle W_{e,q}[\{\Sigma^I\},\{\Omega^J\}] \rangle=-1$} when 
\cred{$e_I=N_I,e_J=0,J\neq I$}, which is indication of fermionic nature of the line/surface operator.
\cblue{Here the time evolution of two pairs of end points (two pairs of black dots) form a closed (invisible undrawn) ribbon that has one side rotating by a $2 \pi$ framing, 
as shown in the Figure 9 of Ref.\cite{Wang1404.7854}, which indicates the spin-statistics (exchange statistics) relation.}
}
\label{fig:BB-self-link}
\end{figure}

Consider the case ${M^{4}}=S^4$. Then integrating out $A^I$ imposes the following condition on $B^I$:
\begin{equation}
	dB^{I}=-\frac{2\pi e_I}{N_I}\,\delta^\perp(\gamma_I),\qquad B^{I}=-\frac{2\pi e_I}{N_I}\,\delta^\perp(\Sigma'_I)
\end{equation} 
where $\Sigma'_I$ is a Seifert surface of $\gamma_I$, that is $\partial {\Sigma'}_I=\gamma_I$.
The effective action is then given by the intersection number of the Seifert surfaces
\begin{equation}
	\int_{S^4}\sum_{IJ}\frac{p_{IJ}N_IN_J}{4\pi N_{IJ}}\,B^I\wedge B^J=
	\sum_{IJ}\frac{\pi p_{IJ}e_I e_J}{ N_{IJ}}\,\#(\Sigma'_I\cap \Sigma'_J).
\end{equation}
While the contribution of the surface operators reads
\begin{multline}
	\sum_{I}q_I\int_{\Omega_I} B^I+\sum_{I,J}\frac{e_Ip_{IJ}N_J}{N_{IJ}}\int_{\Sigma_I}B^{\cred{J}}=\\
	-\sum_I\frac{2\pi e_Iq_I}{N_I}\,\#(\Omega_I \cap \Sigma'_I )
	-\sum_{IJ}\frac{2\pi p_{IJ}e_I e_J}{ N_{IJ}}\,\#(\Sigma_I\cap \Sigma'_J)
\end{multline}
Combining all the terms we get
{
\begin{equation}
	\langle W_{e,q}[\{\Sigma^I\},\{\Omega^J\}] \rangle=
	\prod_{I}\exp\left\{-\frac{2\pi i e_Iq_I}{N_I}\,\text{Lk}(\gamma_I,\Omega_I)\right\}
	\prod_{I,J}\exp\left\{-\frac{\pi ip_{IJ} e_I e_J}{N_{IJ}}\,\#(\Sigma_I\cap\Sigma_J)\right\}
\end{equation}}
where we used that, by the definition of the linking number, \cred{$\#(\Sigma'_I  \cap \Omega_I)=\text{Lk}(\gamma_I,\Omega_I)$} and that $\#((\Sigma_I-\Sigma'_I)\cap (\Sigma_J-\Sigma'_J) )=0$ because intersection number of any two closed surfaces in $S^4$ is zero. Note that the result depends not only on $\gamma_I$, but also on the choice of surfaces $\Sigma_I$ that are bounded by them. This is consistent with the fact that if one changes \cred{$\Sigma_I$ to $\Sigma_I+\delta\Sigma_I$, where $\delta\Sigma_I$} is a closed surface, it is equivalent to changing \cred{$q_I \Omega_I\rightarrow q_I\Omega_I+\sum_J \frac{p_{IJ} e_J N_I}{N_{IJ}}\,\delta\Sigma_J$} in (\ref{B2-surface-op}), and $\delta\Sigma_J$ may have non-trivial linking with $\gamma_J$ (see Fig. \ref{fig:BB-int-number}). Also note that in order to calculate the digonal elements $\#(\Sigma_I\cap \Sigma_I)$ one needs to introduce a framing to $\gamma_I$, that is a choice of a non-zero normal vector along $\gamma_I$. The trivial choice of a generic constant vector leads to $\#(\Sigma_I\cap \Sigma_I)=0$. The example of a framing choice that gives $\#(\Sigma_I\cap \Sigma_I)=1$ is shown in Fig. \ref{fig:BB-self-link}. 

One could also detect the value of $p_{IJ}$ by considering, for example, the partition function of the theory (\ref{BB-action}) on a closed simply-connected spin 4-manifold with the second Betti number $b_2$ and the intersection form $Q_{\alpha\beta}$ on $H^2({M^{4}},\Z)$. Integrating out $A^I$ restricts $N^IB^I/2\pi$ to be a representative of an element from $H^2({M^{4}},\Z)$. Equivalently, 
\begin{equation}
	B^I=\frac{2\pi}{N_I}\sum_{\alpha= 1}^{b_2}n^{I\alpha}\delta^\perp(\Sigma_\alpha)
\end{equation}
where $\Sigma_\alpha$ are representatives of the basis elements of $H^2({M^{4}},\Z)$ and  $n^{I\alpha}\in \Z_{N_I}$ taking into account large gauge transformations. The partition function then reads\footnote{For a generic, not necessarily simply connected, 4-manifold $M^4$ the partition function of a discrete 2-form gauge theory in canonical normalization have the following form:
\begin{equation}
	Z[M^4]=\frac{|H^0(M^4,\prod_i \Z_{N_i})|}{|H^1(M^4,\prod_i \Z_{N_i})|}
	\sum_{b\in H^2(M_4,\prod_i\Z_{N_i})}e^{iS[b]}.
\end{equation} Roughly speaking, the denominator of the normalizaiton factor counts discrete group gauge transformations while the numerator counts ambiguitites in the gauge transformations.}
\begin{equation}
	Z[{M^{4}}]=\prod_{I=1}^s{N_I}\sum_{n^{I\alpha}\in \Z_{N_I}}
	\exp\sum_{I,J,\alpha,\beta}\frac{i\pi p_{IJ}}{N_{IJ}}n^{I\alpha}n^{J\beta}Q_{\alpha\beta}.
	\label{BB-4man}
\end{equation}
Suppose for simplicity that $N_I=N,\forall I=1,\ldots, s$. One can rewrite (\ref{BB-4man}) using Gauss reciprocity formula as follows:
\begin{equation}
	Z[{M^{4}}]=\frac{e^{\frac{i \pi  \sigma(p)\sigma({M^{4}})}{4}}\,N^{3s/2}}{|\det p|^{1/2}}
	\sum_{a\in \Z^{s\cdot b_2}/(p\,\otimes\,Q)\Z^{s\cdot b_2}}
	\exp\left\{-i \pi  N\, p^{IJ}Q^{\alpha\beta}a_{I\alpha}a_{J\beta}\right\}
\end{equation}
where $p^{IJ}$ and $Q^{\alpha\beta}$ are inverse matrices of $p_{IJ}$ and $Q_{\alpha\beta}$, respectively.  Here 
$\sigma(p)$ is the signature of the $p_{IJ}$ matrix, that is the difference between the numbers of positive and the negative eigenvalues of
the matrix. Similarly, $\sigma({M^{4}})$ denotes the signature of ${M^{4}}$, which is by definition is the signature of the intersection \cred{matrix} $Q^{\alpha\beta}$.


\section{Fermionic TQFT/ spin TQFT in 2+1D and 3+1D}

\label{sec:fTQFT}

Now we consider spin-TQFTs which arise from gauging unitary global symmetries of fermionic SPTs (fSPTs). 
We can obtain fermionic discrete gauge spin TQFTs from gauging the $(\Z_2)^n$ symmetry of $\Z_2^f \times (\Z_2)^n$ fSPT.
For example, it is recently known that the 2+1D $ \Z_2^f \times \Z_2$ fSPT, namely the $\Z_2$-Ising-symmetric Topological Superconductor,  
has $\nu \in \mZ_8$ classes \cite{Qi1202.3983, HongYaoRyu1202.5805, GuLevin1304, Neupert2D1403.0953, Morimoto1505.06341}.
\cred{The $\nu$-class of $\Z_2^f \times (\Z_2)^n$ fSPT
is realized by stacking \cgreen{$\nu$ layers of pairs} of chiral and anti-chiral p-wave superconductors ($p+ip$ and $p-ip$), in which boundary supports non-chiral Majorana-Weyl modes.}
Formally, one may interpret this $\mZ_8$ classification from the extended version of group super-cohomology \cite{Gu1201.2648, MCheng1501.01313, WangLinGu1610.08478}
or the cobordism group \cite{Kapustin1406.7329, Freed2016}. 
Yet it remains puzzling what are the \emph{continuum field theories} for these fSPTs and their gauged spin TQFTs, and what are the physical observables that fully characterize them.

Our strategy to tackle this puzzle goes as follows.
In Sec. \ref{sec:Z2fZ2fSPT}, we define fSPT path integrals and its gauged TQFTs for all $\nu \in \mZ_8$ through the cobordism approach in Eq. (\ref{gfSPT-Z2-3D}).
In Sec. \ref{sec:Z2fZ2GSD}, we calculate the GSD on the $T^2$ torus which distinguishes only the odd-$\nu$ from the even-$\nu$ classes. 
In Sec. \ref{sec:ZRP3}, we calculate the path integral $Z[\RP^3]$, a single datum that distinguishes all $\nu\in\mZ_8$ classes.
In Sec. \ref{sec:ST}, we show the $\cT^{xy}$ matrix for the $\Z_2$-gauge flux ('t Hooft line) operator
is another single datum that distinguishes $\nu\in\mZ_8$ classes. By computing the $\cS^{xy}$ and $\cT^{xy}$ matrices,
 we propose our continuum field theories for spin TQFTs and identify their underlying fermionic topological orders through \cite{LanKongWen1507.04673},
 shown in Table \ref{table:Z8-gauged}.  In Sec.\ref{sec:Rokhlin} we propose expression for $\Z_2^f\times\Z_2$ fSPT via Rokhlin invariant.
 In Sec.\ref{sec:more2+1D/3+1DsTQFT}, we study more general fSPTs and corresponding spin TQFTs in 2+1 and 3+1D, and their link invariants.

\subsection{2+1D $ \Z_2^f \times \Z_2$ symmetric fermionic SPTs} \label{sec:Z2fZ2fSPT}
Our first non-trivial examples are the spin-TQFTs gauging the unitary $\Z_2$ part of fSPTs with $\Z_2^f \times \Z_2$ symmetry, where $\Z_2^f$ denotes the fermions number parity symmetry. The mathematical classification of such phases using the spin bordism group:
\begin{equation}
\cred{ \Omega^{3,\text{Spin}}_{\text{tor}}(B\Z_2\cgreen{,U(1)}) \equiv \text{Hom}(\Omega^{\text{Spin}}_{3,\text{tor}}(B\Z_2),U(1)) \cong 
\Omega^\text{Spin}_3(B\Z_2) \cong \mZ_8}
\end{equation} 
appeared in \cite{Kapustin1406.7329}. \cgreen{Note that the last isomorphism is non-canonical and follows from the fact that $\Omega^\text{Spin}_3(B\Z_2)$ contains only torsion elements. }
In particular, for the class $\nu\in \mZ_8$, 
the value of the fSPT partition action on a closed 3-manifold ${M^{3}}$ with a spin structure 
$s \in \text{Spin}({M^{3}})$ 
and the background $\Z_2$ gauge connection $a\in H^1({M^{3}},\Z_2)$ is given by
\begin{equation}
e^{iS[a, s]}=e^{\frac{\pi i\nu}{4}\text{ABK}[\text{PD}(a), \;\; s|_{\text{PD}(a)}]}
\end{equation} 
where $\text{PD}$ stands for the Poincar\'e dual.
The $\text{PD}(a)\subset {M^{3}}$ denotes a (possibly unoriented) surface\footnote{For cohomology with $\Z_2$ coefficients, it is always possible to find a smooth representative of the Poincar\'e dual.} in ${M^{3}}$ representing a class in $H_2({M^{3}},\Z_2)$ Poincar\'e dual to $a\in  H^1({M^{3}},\Z_2)$.
The $s|_{\text{PD}(a)}$ is the $\text{Pin}^-$ structure on $\text{PD}(a)$ obtained by the restriction of $s$, and $\text{ABK}[\ldots]$ denotes $\mZ_8$ valued Arf-Brown-Kervaire inavariant of Pin$^-$ 2-manifold $\text{PD}(a)$ (which is its Pin$^-$ bordism class). Although there is no local realization of Arf-Brown-Kervaire invariant via characteristic classes, \textit{schematically} one can write:
\begin{equation}  \label{Z8-action}
 S[a,s]=\frac{\pi\nu}{4}\int_{{M^{3}}}\,a\cup \text{ABK}.
\end{equation} 
where 
\begin{equation}
 \int_{\Sigma}\text{ABK} \equiv \text{ABK}[\Sigma]
\end{equation} 
for any possibly unoriented surface $\Sigma$ embedded into ${M^{3}}$. The corresponding spin-TQFT partition function reads\footnote{Which can be unterstood as expression of type (\ref{eq:Phi(A)}), that is with $B$ fields already integrated out.}
\begin{equation} 
\label{gfSPT-Z2-3D}
 Z[{M^{3}},s]=\frac{1}{2}\sum_{a\,\in H^1({M^{3}},\Z_2)}e^{iS[a,s]}.
\end{equation} 
Starting from Eq.(\ref{gfSPT-Z2-3D}) we explicitly check that the resulting TQFTs for various values of $\nu \in \mZ_8$ are as described in Table \ref{table:Z8-gauged}.
\begin{table}
	
\footnotesize
    \hspace{-5.05em}\begin{tabular}{ | c | c | c | c | c | c | c |}
    \hline
    $\nu$ & TQFT description (Local action) & $\begin{array}{c}\text{Link}\\ \text{inv.}\end{array}$  & $\GSD_{T^2_\text{o}|T^2_\text{e}}$ & $Z[\RP^3]$ & $\cS^{xy}$ & $\cT^{xy}$   \\ \hline
$0$ & 
$\begin{array}{c}
 \text{level $2$ $BF$ theory $\cong$}\\ 
 \text{level $K=\left(
\begin{array}{cc}
 0 & 2 \\
 2 & 0 \\
\end{array}
\right)$   $U(1)^2$ CS $\cong$}\\
\Z_2\text{-toric code}
\end{array}$
 & Lk 
 &
4b $\mid$ 4b
&
1
 &  $\left(
\begin{array}{cccc}
 \frac{1}{2} & \frac{1}{2} & \frac{1}{2} & \frac{1}{2} \\
 \frac{1}{2} & \frac{1}{2} & -\frac{1}{2} & -\frac{1}{2} \\
 \frac{1}{2} & -\frac{1}{2} & \frac{1}{2} & -\frac{1}{2} \\
 \frac{1}{2} & -\frac{1}{2} & -\frac{1}{2} & \frac{1}{2} \\
\end{array}
\right)$ & $\left(
\begin{array}{cccc}
 1 & 0 & 0 & 0 \\
 0 & 1 & 0 & 0 \\
 0 & 0 & 1 & 0 \\
 0 & 0 & 0 & -1 \\
\end{array}
\right)$ 
\\ \hline    
 $1$ & $\begin{array}{c}
 \text{Ising}\,\times\,{p-ip}\cong \\
\text{Ising}\,\times\,\overline{\text{spin-Ising}}\cong  \\
U(2)_{2,-4}\,\times\,(SO(3)_{-1}\times U(1)_1) \;{\text{CS}}      
       \end{array}$
& Arf  
&
3f $\mid$ 3b
&
$\frac{(1+e^{\pm\frac{\pi i}{4}})}{2}$
& $\left(
\begin{array}{cccc}
 \frac{1}{2} & \frac{1}{2} & \frac{1}{\sqrt{2}}  \\
 \frac{1}{2} & \frac{1}{2} & -\frac{1}{\sqrt{2}}  \\
 \frac{1}{\sqrt{2}} & -\frac{1}{\sqrt{2}} & 0  \\
\end{array}
\right)
$ & $\left(
\begin{array}{cccc}
 1 & 0 & 0  \\
 0 & -1 & 0  \\
 0 & 0 & e^{\frac{\pi i}{8}}
\end{array}
\right)$ 
\\ \hline    
$2$ & level $K=\left(
\begin{array}{cc}
 0 & 2 \\
 2 & -1 \\
\end{array}
\right)$  $U(1)^2$ CS & Lk  
&
4b $\mid$ 4b
&
$\frac{(1+e^{\pm\frac{\pi i2}{4}})}{2}$
& $\left(
\begin{array}{cccc}
 \frac{1}{2} & \frac{1}{2} & \frac{1}{2} & \frac{1}{2} \\
 \frac{1}{2} & \frac{1}{2} & -\frac{1}{2} & -\frac{1}{2} \\
 \frac{1}{2} & -\frac{1}{2} & \frac{i}{2} & -\frac{i}{2} \\
 \frac{1}{2} & -\frac{1}{2} & -\frac{i}{2} & \frac{i}{2} \\
\end{array}
\right)$ & $\left(
\begin{array}{cccc}
 1 & 0 & 0 & 0 \\
 0 & 1 & 0 & 0 \\
 0 & 0 & e^{\frac{i \pi }{4}} & 0 \\
 0 & 0 & 0 & e^{-\frac{3}{4} i \pi } \\
\end{array}
\right)$ 
\\ \hline    
 $3$ & $SU(2)_2\times SO(3)_{-1}$ CS
& Arf 
&
3f $\mid$ 3b
&
$\frac{(1+e^{\pm\frac{\pi i 3}{4}})}{2}$
 & 
 $\left(
\begin{array}{cccc}
 \frac{1}{2} & \frac{1}{2} & \frac{1}{\sqrt{2}}  \\
 \frac{1}{2} & \frac{1}{2} & -\frac{1}{\sqrt{2}}  \\
 \frac{1}{\sqrt{2}} & -\frac{1}{\sqrt{2}} & 0  \\
\end{array}
\right)$  &  $\left(
\begin{array}{cccc}
 1 & 0 & 0  \\
 0 & -1 & 0  \\
 0 & 0 & e^{\frac{3\pi i}{8}}
\end{array}
\right)$  
\\ \hline   
 $4$ &  
 $\begin{array}{c}
 \text{level $K=\left(
\begin{array}{cc}
 0 & 2 \\
 2 & 2 \\
\end{array}
\right)$   $U(1)^2$ CS $\cong$}\\
\Z_2\text{-double semions}
\end{array}$
& Lk
&
4b $\mid$ 4b
&
0
& $\left(
\begin{array}{cccc}
 \frac{1}{2} & \frac{1}{2} & \frac{1}{2} & \frac{1}{2} \\
 \frac{1}{2} & \frac{1}{2} & -\frac{1}{2} & -\frac{1}{2} \\
 \frac{1}{2} & -\frac{1}{2} & -\frac{1}{2} & \frac{1}{2} \\
 \frac{1}{2} & -\frac{1}{2} & \frac{1}{2} & -\frac{1}{2} \\
\end{array}
\right)$ & $\left(
\begin{array}{cccc}
 1 & 0 & 0 & 0 \\
 0 & 1 & 0 & 0 \\
 0 & 0 & -i & 0 \\
 0 & 0 & 0 & i \\
\end{array}
\right)$  
\\ \hline  
 $5$ & $SU(2)_{-2}\times SO(3)_{1}$ CS
 & Arf
 &
3f $\mid$ 3b 
&
$\frac{(1+e^{\pm\frac{\pi i 5}{4}})}{2}$
& $\left(
\begin{array}{cccc}
 \frac{1}{2} & \frac{1}{2} & \frac{1}{\sqrt{2}}  \\
 \frac{1}{2} & \frac{1}{2} & -\frac{1}{\sqrt{2}}  \\
 \frac{1}{\sqrt{2}} & -\frac{1}{\sqrt{2}} & 0  \\
\end{array}
\right)$  &  $\left(
\begin{array}{cccc}
 1 & 0 & 0  \\
 0 & -1 & 0  \\
 0 & 0 & e^{-\frac{3\pi i}{8}}
\end{array}
\right)$ 
\\ \hline 
$6$ & level $K=\left(
\begin{array}{cc}
 0 & 2 \\
 2 & 1 \\
\end{array}
\right)$  $U(1)^2$ CS 
 & Lk 
 &
4b $\mid$ 4b
&
$\frac{(1+e^{\pm\frac{\pi i 6}{4}})}{2}$
&
 $\left(
\begin{array}{cccc}
 \frac{1}{2} & \frac{1}{2} & \frac{1}{2} & \frac{1}{2} \\
 \frac{1}{2} & \frac{1}{2} & -\frac{1}{2} & -\frac{1}{2} \\
 \frac{1}{2} & -\frac{1}{2} & -\frac{i}{2} & \frac{i}{2} \\
 \frac{1}{2} & -\frac{1}{2} & \frac{i}{2} & -\frac{i}{2} \\
\end{array}
\right)$ & $\left(
\begin{array}{cccc}
 1 & 0 & 0 & 0 \\
 0 & 1 & 0 & 0 \\
 0 & 0 & e^{-\frac{i \pi }{4}} & 0 \\
 0 & 0 & 0 & e^{\frac{3}{4} i \pi } \\
\end{array}
\right)$
\\ \hline    
 $7$ & $\begin{array}{c}
 \overline{\text{Ising}}\,\times\,{p+ip}\cong  \\
\overline{\text{Ising}}\,\times\,{\text{spin-Ising}}\cong  \\
U(2)_{-2,4}\,\times\,(SO(3)_{1}\times U(1)_{-1}) \;{\text{CS}}     
       \end{array}$
       & Arf
             &
3f $\mid$ 3b
&
$\frac{(1+e^{\pm\frac{\pi i 7}{4}})}{2}$
 & $\left(
\begin{array}{cccc}
 \frac{1}{2} & \frac{1}{2} & \frac{1}{\sqrt{2}}  \\
 \frac{1}{2} & \frac{1}{2} & -\frac{1}{\sqrt{2}}  \\
 \frac{1}{\sqrt{2}} & -\frac{1}{\sqrt{2}} & 0  \\
\end{array}
\right)$ & $\left(
\begin{array}{cccc}
 1 & 0 & 0  \\
 0 & -1 & 0  \\
 0 & 0 & e^{-\frac{\pi i}{8}}
\end{array}
\right)$ 
\\ \hline    
    \end{tabular}
\caption{Table of spin TQFTs 
as fermionic $\Z_2$ gauge theories
that arise from gauging $\Z_2$ symmetry part of $\Z_2^f\times\Z_2$-fSPTs, for different classes of $\nu\in\mZ_8$ 
with total 8 classes.
The first column of the table shows the $\nu\in\mZ_8$ class. 
The second column shows the continuum TQFTs that we obtain by gauging fSPTs.
We use the description of the Ising TQFT in terms of Chern-Simons theory (CS) 
$U(2)_{2,-4}\cong (SU(2)_2\times U(1)_{-4})/\Z_2$
from \cite{Seiberg:2016rsg}. 
By $SO(3)_1$, we denoted the spin-CS theory (see e.g. \cite{jenquin2006spin}) with 
the level normalized such that the states on $T^2$ are subset of $SU(2)_2$ states corresponding to $SU(2)$ representations with odd dimension (1 and 3). 
The third column (``Link inv.'')  shows the topological invariant through which the expectation value of the system of line operators supported on a link in $S^3$ can be expressed. 
The {fourth} column in the right shows the GSD on $T^2$, in terms of
the spin 2-tori as ${T^2_\text{o}}$ and ${T^2_\text{e}}$ with the odd or even parity. 
The ``b'' stands for boson and the ``f'' for fermions.
%
On one hand, the Ising and the $SU(2)_2$ TQFTs contain 3 \textit{bosonic} (i.e. with $(-1)^F=1$) anyons. 
On the other hand, the spin-Ising and the $SO(3)_1$ spin-CS have 1 bosonic state on $T^2$ for any even spin structure, and 
have 1 fermionic ($(-1)^F=-1$) state for the odd spin structure. 
The theories with $\nu=0\mod 2$ contain 4 bosonic states for any choice of the spin structure.
\cred{The fifth column shows that $Z[\RP^3]$ distinguishes all $\nu\in\mZ_8$ classes.}
The last two columns show
the 
{reduced modular}
$\cS^{xy}$ and $\cT^{xy}$ matrices {(see main text for details)}. 
We can compute $\cS^{xy}$ and $\cT^{xy}$ based on the description Eq.(\ref{gfSPT-Z2-3D}),
and find our data consistent with the spin TQFTs that we associate with in the second column.
{
Our spin TQFTs can be identified 
as fermionic topological orders through \cite{LanKongWen1507.04673},
which denoted as 
$6^F_0$ for odd $\nu$,
and $8^F_0$ or $4^B_0$ for even $\nu$. 
We find the correspondence that 
$\nu=0$ as $4^{B,a}_0$,
$\nu=1$ as $\CF_0 \boxtimes 3^{B}_{1/2}$,
$\nu=2$ as $4^{B}_{1}$,
$\nu=3$ as $\CF_0 \boxtimes 3^{B}_{3/2}$,
$\nu=4$ as $4^{B,b}_0$,
$\nu=5$ as $\CF_0 \boxtimes 3^{B}_{-3/2}$,
$\nu=6$ as $4^{B}_{-1}$,
and
$\nu=7$ as $\CF_0 \boxtimes 3^{B}_{-1/2}$.
}
}
\label{table:Z8-gauged}
\end{table}

\subsubsection{Ground state degeneracy (GSD): Distinguish the odd-$\nu$ and even-$\nu$ classes} \label{sec:Z2fZ2GSD}
The first step in identifying the TQFT is calculating ground state degeneracy on $T^2$. Since we deal with the
 spin-TQFT it is necessary to specify the choice  of spin structure on $T^2$. There are 4 choices corresponding to 
 the periodic (P) or anti-periodic boundary (A) conditions along each of two cycles: (P,P), (A,P), (P,A), (A,A). As we will see the Hilbert space \cred{up to an isomorphism} only depends on the parity (the value \cred{of the Arf invariant in $\mZ_2$}), which is odd for (P,P), and is even for (A,P), (P,A), (A,A). We will denote the corresponding spin 2-tori as ${T^2_\text{o}}$ and ${T^2_\text{e}}$. The GSD can be counted by considering partition function on 
${M^{3}}=T^3=T^2_{\text{e}}\times S^1$ 
or 
$T^2_{\text{o}}\times S^1$ 
where we put either periodic or anti-periodic boundary conditions on the time circle $S^1$. 

We denote their GSD as $\GSD_{T^2_{\text{e}}}$ and $\GSD_{T^2_{\text{o}}}$ respectively.
We find that 
\bea
&&\GSD_{T^2} (\nu=\text{odd}) = 
\left\{\begin{array}{c}
\GSD_{T^2_\text{o}}= 3 \text{   (fermions)}, \\
\GSD_{T^2_\text{e}}= 3 \text{   (bosons)}.
\end{array}\right.\\
&&\GSD_{T^2} (\nu=\text{even}) = 
\left\{\begin{array}{c}
\GSD_{T^2_\text{o}}= 4 \text{   (bosons)}, \\
\GSD_{T^2_\text{e}}= 4 \text{   (bosons)}.
\end{array}\right.
\eea
If we account all possible spin structures,  the odd-$\nu$ theories have 3 bosonic and 3 fermionic states (6 states in total), and 
the even-$\nu$ theories have 4 bosonic states in total.

{We can define $\hat{\CS}^{xy}$ and $\hat{\CT}^{xy}$ as generators of $SL(2,\Z)$, the mapping class group of $T^2$, which permute spin structures as follows:
\begin{equation}
	\hat{\CS}^{xy}:\begin{array}{ccc}
	\text{(P,P)} & \mapsto & \text{(P,P)} \\
	\text{(A,P)} & \mapsto & \text{(P,A)} \\
	\text{(P,A)} & \mapsto & \text{(A,P)} \\
	\text{(A,A)} & \mapsto & \text{(A,A)} 
	\end{array}
	\qquad
	\hat{\CT}^{xy}:\begin{array}{ccc}
	\text{(P,P)} & \mapsto & \text{(P,P)} \\
	\text{(A,P)} & \mapsto & \text{(A,A)} \\
	\text{(P,A)} & \mapsto & \text{(P,A)} \\
	\text{(A,A)} & \mapsto & \text{(A,P)}. 
	\end{array}
\end{equation}
So in general, the corresponding quantum operators act between Hilbert spaces for different spin 2-tori $T^2$. Only the unique odd spin structure (that is (P,P)) is invariant. However in our case the Hilbert space on $T^2$ with spin structure $s$ has form of $\CH_{T^2_s}=\tilde{\CH}_{T^2}\otimes {\CH}^{\text{1-dim}}_s$. Here $\tilde{\CH}_{T^2}$ is an $s$-independent purely bosonic Hilbert space that is 3 (4)-dimensional for odd (even) $\nu$. The  ${\CH}^{\text{1-dim}}_s$ is a \textit{one-dimensional} Hilbert space (of spin-Ising $\cong p+ ip$ superconductor, or spin-$SO(3)_1$ Chern-Simons, or their conjugates). The reduced modular $\cS^{xy}$ and $\cT^{xy}$ matrices in Table \ref{table:Z8-gauged} are representations of $\hat{\CS}^{xy}$ and $\hat{\CT}^{xy}$ elements acting on the reduced Hilbert space $\tilde{\CH}_{T^2}$.}

\subsubsection{$Z[\RP^3]$: Distinguish $\nu\in\mZ_8$ classes} \label{sec:ZRP3}

The easiest way to destinguish different TQFTs with the same number of states (say the 4 states for the odd-$\nu$ and the 3+3 states for the even-$\nu$) is to calculate the partition function on $\RP^3$:
\begin{equation}
 Z[\RP^3]=\frac{1}{2}\sum_{a\in H^1(\cred{\RP^3},\Z_2)\cong \Z_2}e^{\frac{\pi i\nu}{4}\text{ABK}[\text{PD}(a)]}=\frac{1}{2}(1+e^{\frac{\pi i\nu}{4}\text{ABK}[\RP^2\subset \RP^3]})=\frac{1}{2}(1+e^{\pm\frac{\pi i\nu}{4}})
\end{equation} 
 where $\pm$ corresponds to the choice of spin structure on $\RP^3$. 
One compares it with the expression via $\cS^{xy}$ and $\cT^{xy}$ matrices:
\begin{equation}
 Z[\RP^3]=(\cS^{xy} (\cT^{xy})^2 \cS^{xy})_{0,0}
\end{equation} 
based on the $(0,0)$ component of the right hand side matrix.
This gives the precise map between the gauged fSPTs for different values of $\nu \in\mZ_8$ and the known fermionic topological orders \cite{LanKongWen1507.04673} as listed in our Table \ref{table:Z8-gauged}. To summarize, we show that the $Z[\RP^3]$ is one simple single datum that distinguishes $\nu\in\mZ_8$ classes of fSPTs.

\subsubsection{$\cS^{xy}$ and $\cT^{xy}$: The mutual- and self-exchange braiding statistics for $\nu\in\mZ_8$ classes} \label{sec:ST}

\begin{figure}[!h]
\centering
\includegraphics[scale=1]{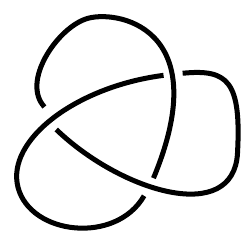}
\caption{Trefoil knot. It has non-trivial Arf invariant (unlike unknot) and can be detected by the \cred{2+1D fermionic (but not bosonic)} gauged SPT with $\Z_2$ symmetry. 
\cred{If the 't Hooft $\Z_2$ flux line (\cgreen{or} the vison in condensed matter terminology), as a worldline of the $\sigma$ anyon, forms \cgreen{a} knot, 
then the trefoil knot gives a statistical Berry phase $(-1)$ for the odd $\nu$ \cgreen{classes}, while it gives a trivial $(+1)$ for the even $\nu$ \cgreen{classes}. 
So the trefoil knot \cgreen{distinguishes odd $\nu$ from even $\nu$} for $\nu \in \mZ_8$ classes of $\Z_2^f\times (\Z_2)^2$ fSPTs.}
}
\label{fig:trefoil}
\end{figure}

Another way is to calculate directly 
the modular data  $\cS^{xy}$ and $\cT^{xy}$ matrices starting from the description Eq.(\ref{gfSPT-Z2-3D}) and computing 
the partition function with line operators supported on the corresponding links. We recall that:
\begin{itemize}
\item A Hopf link for $\cS^{xy}_{mn}$ between two line operators of anyons ($m,n$) encodes the mutual-braiding statistics data between two anyons.
\cgreen{For Abelian anyons, $\cS^{xy}_{mn} \propto e^{\ii \theta_{mn}}$ encodes the Abelian Berry statistical phase $e^{\ii \theta_{mn}}$ of anyons ($m,n$), and, up to an overall factor, is 
related to the total quantum dimensions of all anyons.}

\item A framed unknot for $\cT^{xy}_{nn}$ of a line operator of an anyon ($n$) encodes the self exchange-statistics or equivalently \cred{the spin statistics (also called the topological spin)} of the anyon.  
\end{itemize}
As in the bosonic case, 
the possible nontrivial line operators are the Wilson loop $\exp(\pi i\int_{\gamma'} a)$ and the 't Hooft loop, which imposes 
the condition $da=\delta^\perp(\gamma)$. \cred{Another possibility is a line defect with a non-trivial spin-structure on its complement}. In particular, 
consider the case when we have Wilson loop operators supported on the connected loops $\gamma'_I\subset S^3$, and 't Hooft operators supported on the connected loops 
\cred{$\gamma_J\subset S^3$}.
{
The Wilson loop $\gamma'_I$ is the $\Z_2$-charge loop, while the 't Hooft loop $\gamma_J$ is the $\Z_2$-gauge flux loop (also called the vison loop in condensed matter).}  We consider the loop operators:
\begin{equation}
 W[\{\gamma'_I\},\{\gamma_J\}]=\prod_{I,J} e^{\pi i\int_{\gamma'_J} a}
\delta(da-\sum_{J}\delta^\perp(\gamma_J)),
\end{equation} 
then its expectation value in path integral gives
\begin{equation}
 \langle W[\{\gamma'_I\},\{\gamma_J\}] \rangle \equiv 
\sum_{a\,\in H^1({M^{3}},\Z_2)}e^{iS[a,s]}W[\{\gamma'_I\},\{\gamma_J\}]=
e^{\sum_{I,J}\pi i\,\text{Lk}(\gamma'_I,\gamma_J)}
e^{\frac{\pi i\nu}{4}\text{ABK}(\Sigma)}.
\end{equation} 
Here $\Sigma$ is such that 
$\partial \Sigma= \sqcup _J\gamma_J$ 
and the framing on the link components $\gamma_J$ is induced by $\Sigma$. 
$\text{ABK}(\Sigma)$ is the Arf-Brown-Kervaire invariant of embedded surface with the boundary $\Sigma\subset S^3$ \cite{kirby2004local}. Note that it can be expressed via the Arf invariant of \textit{unframed} link $\{\gamma_J\}$ as follows
\footnote{Note that both ABK and Arf invariant only defined for the \textit{proper} links, that are links such that each component evenly links the rest. It can also be naturally extended for all links, taking values in $\mZ_8^*\equiv \mZ_8 \sqcup \infty$ instead\cite{kirby2004local}, that is $e^{\frac{\pi i}{4}\text{ABK}}=0$ (equivalently, $(-1)^\text{Arf}=0$) for the improper links. This means that $\langle W \rangle=0$ in this case.
}
\footnote{The Arf invariant of a link can be expressed via Arf invariants of individual components\cite{kirby2004local}:
\begin{equation}
	\text{Arf}[\{\gamma_J\}]=\sum_I\text{Arf}[\gamma_I]+
	\frac{1}{4}\sum_{I<J}(\lambda(\gamma_I,\gamma_J)+\text{Lk}(\gamma_I,\gamma_J)+\sum_{I,J,K}\bar{\mu}(\gamma_I,\gamma_J,\gamma_K),
\end{equation}
where $\lambda$ is the Sato-Levine linking invariant and $\bar{\mu}$ is the Milnor triple linking number.
}:
\begin{equation} \label{eq:ABKSigma}
	\text{ABK}[\Sigma]=4\text{Arf}[\{\gamma_J\}]+\frac{1}{2}\sum_{I,J}\text{Lk}(\gamma_I,\gamma_J).
\end{equation}
Therefore we have:
\begin{equation}
	 \langle W[\{\gamma'_I\},\{\gamma_J\}] \rangle 
	=
	(-1)^{\sum_{I,J}\text{Lk}(\gamma'_I,\gamma_J)+\nu\text{Arf}[\{\gamma_J\}]}
	\cdot e^{\frac{\pi i\nu}{8}\sum_{I,J}\text{Lk}(\gamma_I,\gamma_J)}.
	\label{W-arf}
\end{equation}
Note that when $\nu$ is even, the dependence on $\text{Arf}$ invariant goes away, and the expectation value becomes as in Eq.(\ref{Ab-CS-linking}) with the level matrix given in Table \ref{table:Z8-gauged}. Note that \cgreen{the} trefoil \cred{knot} provides an example with non-zero Arf invariant, see Fig. \ref{fig:trefoil}.

Remember that the 't Hooft loop $\gamma_J$ is equivalent to the $\Z_2$-gauge flux vison loop,
where we gauge the $\Z_2$-symmetry of $\Z_2^f \times \Z_2$-fSPT.
{
We anticipate such a  't Hooft loop $\gamma_J$ as the $\Z_2$-gauge flux may be identified with the sigma anyon $\sigma$ \cred{either in 
the Ising TQFT or in the $SU(2)_2$} Chern-Simons (CS) theory.
}
With this in mind, to reproduce the non-trivial element of the $\cT^{xy}$-matrix, 
we can take the link to be a framed unknot of the 't Hooft loop $\gamma_J$. 
Thus we denote $\{\gamma_J\}=\{\circlearrowleft_p\}$. Here the $\circlearrowleft_p$ is an oriented unknot with a framing of any integer $p$.
The framing $p$ means that the line is $2\pi$-twisted by $p$ times, as being Dehn twisted by $p$ times and then glued to a closed line.
We derive that
\begin{equation} \label{eq:WTp}
	\cred{ 
	\langle W[\emptyset,\{\circlearrowleft_p\}] \rangle 
	= e^{\frac{\pi i\nu}{8}p}
	}
\end{equation}
since $\text{Arf}[\circlearrowleft]=0$. 
{
This reproduces an $e^{\frac{\pi i\nu}{8}}$ element of the $\cT^{xy}$-matrix, with a power $p$.
Remember that $\cT^{xy}$ matrix represents the \emph{self-statistics} and equivalently
the \emph{spin} (\cred{also named} as the \emph{topological spin}) of the quasi-particle.
The result $e^{\frac{\pi i\nu}{8}p}$
confirms post-factum our earlier prediction that the 
$\Z_2$-gauge flux 't Hooft line operator should be identified with the same line operator for the sigma anyon $\sigma$.
Thus, we further establish the correspondence between the gauged fSPT Eq.(\ref{gfSPT-Z2-3D})
and the TQFTs in the second column of Table \ref{table:Z8-gauged} for all $\nu \in \mZ_8$ classes.
}

\cred{When $\nu$ is odd}, similarly one can confirm that, \cred{both} Ising and $SU(2)_2$ anyons $\psi$ can be realized by Wilson lines and use the general expression (\ref{W-arf}) to reproduce the other elements\footnote{In order to calculate the elements of $\cS^{xy}$ matrix it is important to fix the normalization of Wilson and flux lines. In particular, the flux line should have an extra $\sqrt{2}$ factor, which is easy to fix by considering a pair of flux lines embedded in the obvious way into $S^2\times S^1$ and requiring that the corresponding path integral is $1$. 
} 
of $\cT^{xy}$ and $\cS^{xy}$. For example the fact that the diagonal element of $\cS^{xy}$ for flux lines is zero follows from the fact that Hopf link is not a proper link and $(-1)^\text{Arf}=0$.


Note that the appearance of the Arf invariant for the odd $\nu$ is consistent with the following two facts: 
(1) The expectation values of Wilson lines supported on a link $\CL$ in the fundamental representation of the $SU(2)_2$ CS theory 
is given by the Jones polynomial of the link $J[\CL]\in \Z[q^{1/2},q^{-1/2}]$ at $q=e^{\frac{2\pi\i}{2+2}}=i$ \cite{Witten:1988hf}
\cgreen{,
where Jones polynomial is an element of $\Z[q^{1/2},q^{-1/2}]$, the space of Laurent polynomials in $q^{1/2}$ with integer coefficients. 
}
(2) The value of the Jones polynomial (up to a simple normalization related factor) at $q=i$ is given by the Arf invariant $J[\CL]|_{q=i}\propto (-1)^{\text{Arf}[\CL]}$  \cite{jones1987hecke,murakami1986}.

To summarize, we show that the modular data $\cS^{xy}$ and $\cT^{xy}$ computed in our Table \ref{table:Z8-gauged}
also distinguish $\nu\in\Z_8$ classes of fSPTs.

\subsubsection{{Fermionic Topological Superconductor and Rokhlin invariant}} \label{sec:Rokhlin}

We explore further on the partition function of $\Z_2^f\times \Z_2$ fSPTs, namely the $\Z_2$-symmetric fermionic Topological Superconductors. 
First let us notice that the partition function of $SU(2)_2$ Chern-Simons theory (CS) on a closed 3-manifold can be expressed via Rokhlin invariant \cite{kirby19913}:
\begin{equation}
	Z_{SU(2)_2}[{M^{3}}]=\frac{1}{2}\sum_{s\in \text{Spin}({M^{3}})}e^{-\frac{3\pi i}{8}\mu({M^{3}},s)}
\end{equation}
where the Rokhlin invariant  $\mu ({M^{3}},s)$ of a 3-manifold ${M^{3}}$ equipped with the spin-structure $s\in \text{Spin}({M^{3}})$ is {defined} as:
\begin{equation}
	\mu ({M^{3}},s)=\sigma({M^{4}})\mod 16.
\end{equation}
The $\sigma({M^{4}})$ is the signature of any spin 4-manifold bounded by ${M^{3}}$ so that spin structure on ${M^{3}}$ is induced by spin structure on ${M^{4}}$. Similarly, for its spin version, that is the $SO(3)_1$ spin CS:
\begin{equation}
		Z_{SO(3)_1}[{M^{3}},s]=e^{-\frac{3\pi i}{8}\mu({M^{3}},s)}.	
\end{equation}
Combining those together, we have:
\begin{multline}
	Z_{SU(2)_2\times SO(3)_{-1}}[{M^{3}},s]=\frac{1}{2}\sum_{s'\in \text{Spin}({M^{3}})}e^{-\frac{3\pi i}{8}\mu({M^{3}},s')+\frac{3\pi i}{8}\mu({M^{3}},s)}=\\ \frac{1}{2}\sum_{a\in H^1({M^{3}},\Z_2)} 
	e^{-\frac{3\pi i}{8}(\mu({M^{3}},s+a)-\mu({M^{3}},s))}
\end{multline}
where we used the fact that the spin structures form an affine space over $H^1({M^{3}},\Z_2)$. Comparing it with (\ref{gfSPT-Z2-3D}) 
at $\nu=3$ for the ${SU(2)_2\times SO(3)_{-1}}$ Chern-Simons theory,
this suggests that\footnote{This should follow from the following formula \cite{guillou62extension}
\begin{equation}
	\mu({M^{3}},s)=\sigma({M^{4}})-(\text{PD}(w_2)\cdot\text{PD}(w_2))+2\text{ABK}[\text{PD}(w_2)]
\end{equation}
where ${M^{4}}$ is \textit{any} (not necessarily spin) 4-manifold bounded by ${M^{3}}$, and $w_2$ is \cgreen{the} \cred{relative} Stiefel-Whitney class.
}
\begin{equation} \label{eq:RokhlinNew}
	\mu({M^{3}},s)-\mu({M^{3}},s+a) = 2\text{ABK}[\text{PD}(a),s|_{\text{PD}(a)}] \mod 16
\end{equation}
and that the partition function of fSPT is given by
\begin{equation} \label{eq:RokhlinNew2}
	Z_{\text{fSPT}_\nu}[{M^{3}},s,a]=e^{\frac{\pi i\nu}{8}(\mu({M^{3}},s)-\mu({M^{3}},s+a))}.
\end{equation}
The fact that it is a cobordism invariant can be understood as follows. 
Let ${M^{3}}$ and ${M^{3}}'$ be 3-manifolds equipped with spin-structures $s,s'$ and $\Z_2$ gauge fields $a,a'$ (that is $\Z_2$ principal bundles over ${M^{3}}$ and ${M^{3}}'$, or equivalently maps ${M^{3}},{M^{3}}'\rightarrow B\Z_2$ ). 
Suppose 
these spin 3-manifolds with $\Z_2$ gauge bundles 
 represent the same class in $\Omega^\text{Spin}_3(B\Z_2)$.
 Then there exists a 4-manifold ${M^{4}}$ equipped with spin structure $s_4$ and $\Z_2$ gauge field $a_4\in H^1({M^{4}},\Z_2)$ such that $\partial {M^{4}}={M^{3}}' \sqcup (-{M^{3}})$ and $s_4|_{{M^{3}}}=s$, $s_4|_{{M^{3}}'}=s'$, $a_4|_{{M^{3}}}=a$, $a_4|_{{M^{3}}'}=a'$. It follows that $(s_4+a_4)|_{{M^{3}}}=s+a$, $(s_4+a_4)|_{{M^{3}}'}=s'+a'$. Therefore, by the definition of Rokhlin invariant
\begin{equation}
	\mu({M^{3}},s)-\mu({M^{3}}',s')=\sigma({M^{4}}) \mod 16
\end{equation}
and
\begin{equation}
	\mu({M^{3}},s+a)-\mu({M^{3}}',s'+a')=\sigma({M^{4}}) \mod 16,
\end{equation}
and therefore
\begin{equation}
	Z_{\text{fSPT}_\nu}[{M^{3}},s,a]=Z_{\text{fSPT}_\nu}[{M^{3}}',s',a'].
\end{equation}

\subsection{Other examples of 2+1D/3+1D spin-TQFTs and $\Z_2^f\times (\Z_2)^n$ fermionic SPTs: \\
Sato-Levine invariant and more} \label{sec:more2+1D/3+1DsTQFT}

In Table \ref{table:fSPT-gauged}, we propose other examples of spin-TQFTs with action (formally written) similar to (\ref{Z8-action}). The idea is that if we have a collection of $\Z_2$ gauge fields $a_i\in H^1({M^{d}},\Z_2)$, there is lesser-dimensional fPST with time-reversal symmetry (with $T^2=1$) living on the intersection of domain walls (Poincar\'e dual to $a_i$, which is, in general, non-orientable). The 1-cocycle $\eta$ is \textit{formally} defined such that $(-1)^{\int_{S^1}\eta}=\pm 1$ depending on the choice of spin structure on $S^1$. It can be interpreted as the action of the non-trivial 0+1D fSPT with no unitary global symmetry, that is the theory of one free fermion (the 0+1D fSPT partition function is $1$ for the choice of \cred{anti-periodic} boundary conditions on the fermion, and $-1$ for periodic boundary conditions). 

The corresponding link invariants are similar to those appeared in bosonic theories, but instead of counting points of intersection between loops/surfaces/Seifert (hyper)surfaces, we count $\Omega^{\text{Pin}^-}_{1,2}(\pt)$ bordism classes of 1- and 2-manifolds that appear in the intersection. 
In math literature, the result of such counting sometimes referred as ``framed intersection''. 
Note that in one dimension, 
the $\text{Pin}^-$ bordism group is isomorphic to the stable framed bordism group: $\Omega^{\text{Pin}^-}_1\cong\Omega^{\text{Spin}}_1\cong \Omega^{\text{fr}}_1\cong\pi_1^s\cong \mZ_2$ which, in turn, by Pontryagin-Thom construction is isomorphic to the stable homotopy group of spheres. The $\text{Pin}^-$ structure on the intersection is induced by Spin structure on the ambient space together with the framing on its normal bundle given by vectors which are tangent to the intersecting surfaces \cite{kirby1990pin}. 

\begin{table}[t!]
\footnotesize
\begin{center}
    \begin{tabular}{| c | c|c | c | }
    \hline
   Dim & Symmetry & $\begin{array}{c} \text{Action} \\ (\text{Formal notation}) \end{array}$  &  $\begin{array}{c}\text{Link}\\ \text{invariant}\end{array}$ \\ \hline\hline
2+1D & $\Z_2^f\times \Z_2$ & 
$\frac{\pi}{4}\int a\cup \text{ABK}$ & Arf invariant of a knot/link
\\ \hline   
2+1D & $\Z_2^f\times (\Z_2)^2$ &
${\pi}\int a_1\cup a_2 \cup \eta$
 & $\begin{array}{c}
\text{Sato-Levine invariant\,\cite{sato1984cobordisms}}:\\
 \gamma_1,\gamma_2\mapsto\text{Framed bordism class} \\
  \text{[}\Sigma_1 \cap \Sigma_2 \text{]} \in \pi_1^s\cong \mZ_2
 \end{array}$
 \\ \hline  \hline  
3+1D & $\Z_2^f\times (\Z_2)^2$ (?)& 
$\frac{\pi}{4}\int a_1\cup a_2 \cup \text{ABK}$
& $\begin{array}{c} \Sigma_1,\Sigma_2 \mapsto \\
\text{ABK}[\CV_1 \cap \CV_2] \;\;\;
(\partial\CV_I=\Sigma_I)
\end{array}$
\\ \hline    
3+1D & $\Z_2^f\times (\Z_2)^2$ (?)& 
$\pi\int a_1\cup a_2 \cup a_2\cup \eta$
& 
 $\begin{array}{c}
 \text{Dlk}[\Sigma_1,\Sigma_2]=\\
\text{Framed bordism class} \\
 \text{[}\Sigma_1 \cap \CV_2 \text{]} \in \pi_1^s\cong \mZ_2
 \;\;\text{\cite{sato1984cobordisms, carter2008link}}
\end{array}$
\\ \hline   
3+1D & $\Z_2^f\times (\Z_2)^3$ & 
$\pi\int a_1\cup a_2 \cup a_3\cup \eta$
& $\begin{array}{c}
\Sigma_1,\Sigma_2,\Sigma_3\mapsto\text{Framed bordsim class} \\
 \text{[}\CV_1 \cap \CV_2 \cap \CV_3\text{]}+\cred{[\ldots]} \\
\in \pi_1^s\cong \mZ_2
\end{array}$
\\ \hline   
    \end{tabular}
    \end{center}
\caption{Table of spin TQFTs and the corresponding link invariants.
{As before $\text{ABK}(\Sigma)$ is the Arf(-Brown-Kervaire invariant) of Spin (Pin$^-$) surface $\Sigma$.
The 1-cocycle $\eta$ is \textit{formally} defined by the rule $(-1)^{\int_{S^1}\eta}=\pm 1$  with the dependence on the spin structure choice on $S^1$. }
Notation $\pi_1^s (\cong \Omega_1^{\text{Pin}^-}(\pt))$ stands for the stable framed bordism group of 1-manifolds.
Here $\text{Dlk}[\Sigma_1,\Sigma_2]$ stands for the double linking (Dlk) of two surfaces $\Sigma_1$ and $\Sigma_2$, given
in Ref.\cite{sato1984cobordisms, carter2008link}.
It can detect, for example, two linked surfaces obtained as a twisted spun of the Hopf link. 
Here
the ${[\ldots]}$ means extra terms similar to the ones in Eq.(\ref{triple-linking}) that insure invariance under the choice of three Seifert hypersurfaces.
Note that from the point of view of bordism classification of fSPTs, 
it is not surprising that the emerging link invariants are \cred{cobordism invariants} of links.
{Here the formal actions in Table \ref{table:fSPT-gauged} are consistent topological invariants that we merely propose them 
to detect potential SPT states. However,
there are further obstructions \cite{KapustinThorngren1701.08264,WangGu1703.10937} in order to define SPTs on any closed manifold
through these topological invariants.
Indeed some topological invariants (e.g. the third and fourth rows marked with ``?'') above do not correspond to any SPTs, where we report 
 the details in a companion work \cite{to-appear}.
} }
\label{table:fSPT-gauged}
\end{table}

To clarify, consider for example the case with action $\pi \int a_1\cup a_2\cup \eta$ (second line in Table \ref{table:fSPT-gauged}). 
More precisely, the fSPT partition function on a 3-manifold ${M^{3}}$ is given by
\begin{equation}
e^{iS[a_1,a_2]}=(-1)^{\int_{\text{PD}(a_1)\cap\text{PD}(a_2)}\eta}\equiv 
\prod_{\text{circles in } \text{PD}(a_1)\cap\text{PD}(a_2)}\left\{\begin{array}{cl}
+1, & \text{odd},\\
-1, & \text{even}, 
\end{array}
\text{ spin structure on } S^1
\right. 
\end{equation}
\cred{The circle with \cgreen{the} anti-periodic boundary condition on fermions is spin-bordant to an empty set, 
since it can be realized as a boundary of \cgreen{a} disk. So \cgreen{the partition function of the non-trivial $0+1$D fSPT has value 1 for it}. 
The circle with \cgreen{the} periodic boundary condition forms \cgreen{the} generator of the spin-bordism group, and \cgreen{the partition function of the non-trivial fSPT} has value $-1$.}
Note that induced spin-structures on circles embedded into ${M^{3}}$ can be understood as their framings (trivialization of the tangent bundle) modulo two \cite{kirby1990pin}. If one chooses framing on ${M^{3}}$ compatible with spin structure, the framing on the circle that appears at the intersection of two surfaces is then determined by the two normal vectors tangent to the surfaces. Intuitively the framing on $S^1$ is given by how many times the ``cross'' of intersecting surfaces winds when one goes around the loop. 
Physically this 2+1D fSPT can be constructed by putting a non-trivial 0+1D fSPT (which are classified by $\mZ_2$) state, that is a state with one fermion, on the intersections of pairs domain walls for discrete gauge fields $a_{1,2}\in H^1({M^{3}},\Z_2)$. The partition function of the corresponding spin-TQFT then can be written as follows:
\begin{equation}
Z=\frac{1}{4} \sum_{a_1,a_2\in H^1({M^{3}},\Z_2)}e^{iS[a_1,a_2]}
\end{equation}

\begin{figure}[!h]
\centering
\includegraphics[scale=1]{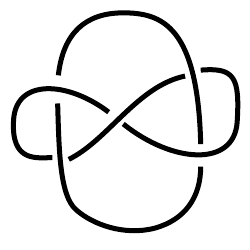}
\caption{Whitehead link of two worldlines $\gamma_1$ and $\gamma_2$. It has a trivial linking number, but non-trivial Sato-Levine invariant $\mathrm{SL}(\gamma_1,\gamma_2)=1$. 
Therefore it is detected by \cred{the 2+1D fermionic (but not bosonic)} gauged SPT with $\Z_2\times \Z_2$ symmetry. 
The two $\Z_2$-flux lines of distinct $\Z_2$ form two components of the link.
\cred{Moreover, the classification of $\Z_2^f\times (\Z_2)^2$ fSPTs shows \cgreen{$(\mZ_8)^2 \times \mZ_4$ distinct} classes \cite{WangLinGu1610.08478}.
In this case, we claim that the Whitehead link can detect \cgreen{odd $\nu\in \mZ_4$ classes} in $\mZ_4$ sub-classes with a Berry phase $(-1)$.}}
\label{fig:whitehead}
\end{figure}

Now let us consider flux line operators ($W[\gamma_I]\propto \,\delta(da_I-\delta^\perp(\gamma_I))$) in this theory supported on a two-component semi-boundary link $\{\gamma_1,\gamma_2\}$ in $S^3$. 
``Semi-boundary'' link is by definition a link for which one can choose Seifert surfaces $\Sigma_I$ (\cred{which satisfies $\partial \Sigma_I =\gamma_I $}) such that $\Sigma_I\cap \gamma_J=0$ for $I\neq J$. 
\cgreen{Note that semi-boundary links should be distinguished from boundary links, the links  that satisfy a stronger condition $\Sigma_I \cap \Sigma_J = 0$ for $I\neq J$.}
Then\footnote{The extra $\exp(\ldots)$ factors are gauge-invariant completions of line operators, similar to the ones in Sec. \ref{sec:AAdA}.}
\begin{equation}
W[\gamma_1,\gamma_2]=\prod_I \delta(da_I-\gamma_I)
e^{\pi i\sum_J\epsilon^{IJ} \int_{\Sigma_I}a^J\cup \eta }
\end{equation}
\begin{equation}
\langle W[\gamma_1,\gamma_2]\rangle
=e^{\pi i \int_{\Sigma_1\cap \Sigma_2 }\eta}=
\prod_{\text{circles in } \Sigma_1\cap \Sigma_2} (-1)^\text{framing}\equiv (-1)^{\mathrm{S}(\gamma_1,\gamma_2)}
\end{equation}
where the framing on the connected components of $\Sigma_1\cap \Sigma_2\subset S^3$ is determined by the normal vectors in the direction of $\Sigma_{1,2}$. This invariant 
$\mathrm{SL}(\gamma_1,\gamma_2)\in\pi^s_1\cong\mZ_2$ 
is known as the (stable) Sato-Levine linking invariant of semi-boundary link \cite{sato1984cobordisms}. It can be used to detect some non-trivial links for which the usual linking number is zero. 
The simplest example of a 2-component link with $\text{Lk}(\gamma_1,\gamma_2)=0$ but 
$\mathrm{SL}(\gamma_1,\gamma_2)=1$ is the Whitehead link (see e.g. \cite{cochran1985geometric}) shown in Fig \ref{fig:whitehead}).

\section{Conclusion}
\label{sec:conclude}

Some final remarks and promising directions are in order:
\begin{enumerate}
\item 
We formulate the continuum TQFTs to verify several statistical Berry phases raised from particle and string braiding processes 
via the 2+1D and 3+1D spacetime path integral formalism (See \cite{1602.05951} and References therein). 
We find the agreement with \cite{1602.05951} which uses a different approach based on surgery theory and quantum mechanics. 
As far as we are concerned, all the TQFTs discussed in our Table \ref{table:TQFTlink}, \ref{table:Z8-gauged} and \ref{table:fSPT-gauged}
can be obtained from dynamically gauging the unitary global symmetries of certain SPTs.			
We also derive the corresponding new link invariants directly through TQFTs.

\item 
{We resolve the puzzle of continuum spin TQFTs that arise from gauging the $\Z_2^f\times \Z_2$-fSPT 
(namely the $\Z_2$-symmetric fermionic Topological Superconductor) partition function (\ref{gfSPT-Z2-3D}) and listed in Table \ref{table:Z8-gauged}}.
In addition, we may also understand the spin bordism group $\Omega^\text{Spin}_3(B\Z_2)\cong \mZ_8$ classification \cite{Kapustin1406.7329} of those fSPT  
in terms of three layers with
three distinct $\mZ_2$ generators respectively through the extended group super-cohomology approach \cite{Gu1201.2648, MCheng1501.01313, WangLinGu1610.08478}. 
\cred{Comparing the cohomology group classification\cite{MCheng1501.01313, WangLinGu1610.08478} \cgreen{with a group $G=\Z_2$
and} Table \ref{table:Z8-gauged},  we can deduce that 
the first $H^3(G,U(1))=\mZ_2$ group generates the bosonic Abelian CS theories for $\nu=0,4$,
the second $H^2(G,\Z_2)=\mZ_2$ group generates the fermionic Abelian spin CS theories for $\nu=2, 6$, and
the third $H^1(G,\Z_2)=\mZ_2$ group generates fermionic non-Abelian spin TQFTs for $\nu=1,3,5,7$.
}

\item Following the above comment,  for the odd-$\nu$ non-Abelian spin TQFTs in Table \ref{table:Z8-gauged}, 
we see that they are formed by a product of a bosonic and a spin TQFT sectors, each with opposite chiral central charges $c_-$.
The Ising TQFT and the $SU(2)_2$ CS both have anyon contents $\{1, \sigma, \psi\}$, while the $p+ip$ superconductor and the $SO(3)_1$ CS
have contents $\{1, f\}$. 
Here we identify the $\sigma$ as the anyon created by the 't Hooft loop of $\Z_2$-gauge flux vison through our $\cT^{xy}$ matrix Eq.(\ref{eq:WTp}).
The $\psi$ is the Bogoliubov fermion, and $f$ is a  \textit{fundamental} fermion.
\cred{The $\sigma$ is a non-Abelian anyon created from the $\Z_2$-gauge flux. \cgreen{We can identify $\sigma$ with a Majorana zero mode} trapped at a half-quantum vortex of a dynamically gauged chiral $p$-wave superconductor \cite{Ivanov20005069}.
}

To recap, we should regard the $\nu=1$ class 
$\Z_2$-symmetric Topological Superconductor
as stacking a $p+ip$ superconductor and a $p-ip$ superconductor together.
More generally, we should regard the $\nu$ class fSPT as stacking $\nu$ copies of $p+ip$ superconductors and  $p-ip$ superconductors.
To obtain the spin TQFTs by gauging the fSPTs, naively we dynamically gauging the superconductor vortices in one sector of theory.
More precisely, $f$ is realized by equipping non-trivial spin structure on the complement of the loop.
The theory of $\{1, \sigma, \psi\}$ is related to $\{1, f\}$ by dynamically gauging the $\Z_2$-flux, thus summing over the spin structures.
The $\Z_2$-gauge flux traps the Majorana mode identified as the $\sigma$ anyon.  
The pair of chiral central charges $c_-$ for two sectors of odd-$\nu$ theories are 
$(\frac{1}{2},-\frac{1}{2})$ for $\nu=1$, 
$(\frac{3}{2},-\frac{3}{2})$ for $\nu=3$,
$(-\frac{3}{2},\frac{3}{2})$ for $\nu=5$, 
$(-\frac{1}{2},\frac{1}{2})$ for $\nu=7$, which satisfy a relation of $(\frac{\nu}{2} ,-\frac{\nu}{2})$ up to mod 4.
\cred{So the total chiral central charges is $c_-=0$ for each theory, as it supposed to be \cgreen{for a} gauged fSPTs.}


\item Comment on \emph{gauging and bosonization}: 
\cred{We like to make a further remark on the relation between 
 SPTs and TQFTs.}
 
By \emph{gauging}, one may mean to probe SPTs by coupling the global symmetry to a background classical probe field.
However, here in our context, in order to convert (short-range entangled) SPTs to (long-range entangled) TQFTs,
we should further make the gauge field dynamical. Namely, in the field theory context, 
\cgreen{this means} summing over all the classical gauge field configuration to define the path integral.
This step is straightforward for bosonic theories in Sec.~\ref{sec:BdA}-\ref{sec:BB-theory}.

\cred{
By \emph{bosonization}, we mean summing over all the spin structures for a fermionic theory \cgreen{which turns} it into a bosonic theory.}

\cred{
It is worthwhile to clarify the physics of \emph{gauging} procedure \cgreen{and its relation to } \emph{bosonization} \cgreen{of} the fermionic theories in Sec.~\ref{sec:fTQFT}.}
\cgreen{The background $\Z_2$ gauge field of fSPTs in Sec.~\ref{sec:fTQFT}'s {Table \ref{table:Z8-gauged}}
can be understood as the difference between spin structures for chiral and anti-chiral factors (i.e. $p+ip$ and $p-ip$ superconductors for $\nu=1$ {of Table \ref{table:Z8-gauged}}). By fixing a the spin structure for one of the chiral factors (say, $p-ip$), the summation over the background gauge field becomes equivalent to the summation over the spin structure for the other factor.
In particular, in the $\nu=1$ case, gauging produces $\text{Ising}\,\times\,{p-ip}$ fTQFT or equivalently 
$\text{Ising}\,\times\,\overline{\text{spin-Ising}}$ fTQFT. The case of other $\nu\in \Z_8$ classes is similar.
}

\item Comment on \emph{the quantum dimension and statistical Berry phase/matrix of anyonic particles/strings}: 
 \cred{In order to discuss the quantum dimensions $d_J$ of loop/surface
operators, one should properly normalize the line/surface operators.
There are different choices of possible normalizations.
\cgreen{In particular, for condensed matter/quantum information literature, 
the common normalization of line operators is such} that
insertion of the pair of operator and anti-operator along a
non-trivial loop in \cgreen{$S^d\times S^1$} in $d+1$D, where $d+1$ is the spacetime dimension,
yields 1. For example, from Sec.~\ref{sec:BdA}-\ref{sec:BB-theory},
it can be shown that quantum dimensions $d_J$ for $\int BdA+A^3$ and
$\int BdA+A^4$ theories are non-Abelian in the sense that $d_J=N$ for some of
the operators that contain $B_J$ fields \cgreen{(where $B_J$ is a 1-form field for $\int BdA+A^3$ theory and
a 2-form field for $\int BdA+A^4$ theory)
in twisted Dijkgraaf-Witten theories with
$G=(\Z_N)^{d+1}$ with $N \geq 2$}.
The non-Abelian nature can be seen clearly as follows:  On a spatial $S^d$ sphere
with a number $\text{N}_{\text{insert}}$  of insertions of the non-Abelian excitations,
the GSD grows as $\GSD \approx (d_J)^{\text{N}_{\text{insert}}}=N^{\text{N}_{\text{insert}}}$.
Thus, the Hilbert space of degenerate ground state sectors has dimension of order \cgreen{$N^{\text{N}_{\text{insert}}}$}}.
 \cred{\cgreen{One} can verify that the quantum dimensions $d_J=N$ is consistent with an independent derivation in Ref.~\cite{Wang1404.7854} based on quantum algebra (without using field theory).}
 
 {In general, 
the spacetime braiding process of anyonic particle/string excitations, on a spatial sphere,
in such a highly degenerate ground state Hilbert subspace (of a dimension of an order 
 $\GSD \approx (d_J)^{\text{N}_{\text{insert}}}$),
would evolve the original state vector
with an additional statistical unitary Berry matrix (thus, non-Abelian statistics).
}

 Normally, the non-Abelian statistics for the non-Abelian anyons and non-Abelian strings 
for these theories are more difficult to compute, because the non-Abelian statistics require a matrix to characterize the changes of ground-state sectors.
Yet for $\int BdA+A^3$ of Sec.~\ref{sec:aaa-theory} and $\int BdA+A^4$ theories of Sec.~\ref{sec:A4-theory}, 
we are able to use the Milnor's triple linking and the quadruple-linking numbers of surfaces to characterize their non-Abelian statistics, thus we boil down
the non-Abelian statistics data to a single numeric invariant.\footnote{The physics here is similar to 
the observation made in \cite{CWangMLevin1412.1781}, although in our case we have obtained more general link invariants that encode all possible nontrivial braiding processes, instead of particular few braiding processes in \cite{CWangMLevin1412.1781}.} 

 \cred{We would also like to point out that the main focus
of \cgreen{the current work} is to derive the more subtle statistical Berry
phases of anyonic particles/strings. For this purpose, we choose a
convenient but different normalization of line/surface operators (result shown in Table \ref{table:TQFTlink}). One can easily modify our prescription above to
encode the \emph{quantum dimension} data.}

\item We remark that the $\int BdA+A^3$ studied in Sec.~\ref{sec:aaa-theory} has been found in \cite{Ferrari0210100, Gu:2015lfa}
that it can be embedded into a non-Abelian Chern-Simons theory 
\begin{eqnarray}
\label{eq:L-general}
S={\frac{1}{4\pi} }\int d^3 x \epsilon^{\mu\nu\rho} \mathcal{K}^{G}_{a \alpha'} \Big(  \mathcal{A}^a_{\mu}(x) \partial_\nu \mathcal{A}^{\alpha'}_{\rho}(x) 
+\frac{1}{3} f_{bc}{}^{a} \mathcal{A}^{\alpha'}_{\mu}(x) \mathcal{A}^b_{\nu}(x) \mathcal{A}^c_{\rho}(x)\Big),
\end{eqnarray}
here $a, b, c, \alpha, \alpha'=1,\dots,6$, with a 6-dimensional Lie algebra. 
We write
\begin{eqnarray}
&&\mathcal{A}_\mu^\alpha T^\alpha \equiv A_\mu^I  X_I + B_\mu^I H_I^*.\\
&& (\mathcal{A}_\mu^1 T^1,\mathcal{A}_\mu^2 T^2, \mathcal{A}_\mu^3 T^3)=
(A_\mu^1  X_1, A_\mu^2  X_2, A_\mu^3 X_3), \nonumber \\
&& ( \mathcal{A}_\mu^4 T^4, \mathcal{A}_\mu^5 T^5, \mathcal{A}_\mu^6 T^6)=
(B_\mu^1  H_1^*,B_\mu^2  H_2^*, B_\mu^3 H_3^*). \nonumber
\end{eqnarray}
Here $\alpha=1,\dots,6$ and $I=1,\dots,3$,
the generic Lie algebra is called the symmetric self-dual Lie algebra \cite{FigueroaO'Farrill:1995cy}, 
such that the generators $H_I^*$ and $X_I$ in particular obey
\begin{align} \label{eq:Lie_algebra}
[H_I^*,H_J^*]=[H_I^*,X_J]=0; \quad [X_I,X_J]=C_{IJK}H_K^*,
\end{align}
where $C_{IJK}$ serves as an appropriate structure constant now. More generally, we simply write
the whole 6-dimensional Lie algebra as $[T_a,T_b]= f_{ab}{}^c T_c$.
Here even if our Killing form $\kappa_{ab}=\kappa(T_a,T_b)=-\Tr(T_a, T_b)$ is degenerate,
as long as we can define a symmetric bilinear form $({\mathcal{K}_{}^{G}})_{IJ}$, 
the $({\mathcal{K}_{}^{G}})_{IJ}$ can replace the degenerate Killing form
to define the Chern-Simons theory Eq.(\ref{eq:L-general}). See Ref. \cite{Gu:2015lfa}'s Sec.~X and Appendix C for further details.

As an example, when $\int BdA+A^3$ in Sec.~\ref{sec:aaa-theory} has a $G=(\Z_2)^3$, 
it can be shown that it is equivalent to a non-Abelian $D_4$ (of a group order 8)  
discrete gauge theory \cite{deWildPropitius:1995cf,Wang1404.7854, Gu:2015lfa, He1608.05393}.

\item 
Surprisingly, non-Abelian TQFTs can be obtained from gauging the Abelian global symmetry of some finite Abelian unitary symmetry group SPTs
, in the sense that the braiding statistics of excitations have non-Abelian statistics.
Non-Abelian statistics mean that the ground state vector in the Hilbert space can change its sector to a different vector, even if
the initial and final configurations after the braiding process are the same. 
This happens for the $\int BdA+A^3$ in 2+1D of Sec.\ref{sec:aaa-theory},
and $\int BdA+A^4$ in 3+1D of Sec.\ref{sec:A4-theory}.
Other examples are the 2+1D/3+1D spin-TQFTs obtained by gauging $\Z_2^f\times (\Z_2)^n$ fSPTs, we also obtain some non-Abelian spin TQFTs.

\item \emph{New mathematical invariants}: 
The \textit{quadruple linking number} of four surfaces
defined in Eq. (\ref{quadruple-surface-linking}),
$\text{Qlk}(\Sigma_1,\Sigma_2,\Sigma_3,\Sigma_4)$,
seems to be a new link invariant that has not been explored in the mathematics literature. 
In Eqs. (\ref{eq:RokhlinNew}-\ref{eq:RokhlinNew2}), we propose a novel reazliation of $\Z_2^f\times \Z_2$ fSPT partition function (previously realized via Arf-Brown-Kervaire invariant in \cite{Kapustin1406.7329}) and Rokhlin invariant.
\end{enumerate}

\section{Acknowledgements}

JW thanks Zhengcheng Gu, Tian Lan, 
Nathan Seiberg, Clifford Taubes and Edward Witten for conversations.
PP gratefully acknowledges the support from Marvin L. Goldberger Fellowship and the DOE Grant DE-SC0009988. 
JW gratefully acknowledges the Corning Glass Works Foundation Fellowship and NSF Grant PHY-1606531.
%
JW's work was performed in part at the Aspen Center for Physics, which is supported by National Science Foundation grant PHY-1066293.
This work is supported by the NSF Grant PHY- 1306313, PHY-0937443, DMS-1308244, DMS-0804454, DMS-1159412 and Center for Mathematical Sciences and Applications at Harvard University.

\bibliographystyle{JHEP}
\bibliography{SPTgb_ref_new,all,mybib} 


\end{document}